\documentclass[final,5p,times,twocolumn,authoryear]{elsarticle}
\usepackage[utf8]{inputenc}
\usepackage{tensor}
\usepackage{hyperref}
\usepackage{amsfonts}
\usepackage{amsmath}

\usepackage{amssymb}
\usepackage{graphicx}%
\usepackage{bigints}
\usepackage{graphicx}
\usepackage{subfigure}
\usepackage{epsf}
\usepackage{bm}
\usepackage{amsmath}
\usepackage{amsfonts}
\usepackage{amssymb}
\usepackage{graphicx}
\usepackage{tabularx}
\usepackage{color}%
\usepackage{multirow}
\providecommand{\U}[1]{\protect\rule{.1in}{.1in}}
\usepackage{bigints}
\newcommand{\be}{\begin{equation}}
\newcommand{\ee}{\end{equation}}
\newcommand{\Be}{\begin{eqnarray}}
\newcommand{\Ee}{\end{eqnarray}}

\newcommand{\mincir}{\raise
-3.truept\hbox{\rlap{\hbox{$\sim$}}\raise4.truept\hbox{$<$}\ }}
\newcommand{\magcir}{\raise
-3.truept\hbox{\rlap{\hbox{$\sim$}}\raise4.truept\hbox{$>$}\ }}

\newcolumntype{Y}{>{\centering\arraybackslash}X}
\providecommand{\U}[1]{\protect\rule{.1in}{.1in}}

\usepackage[utf8]{inputenc}

\journal{Physics of the Dark Universe}

\begin{document}

\journal{Physics of the Dark Universe}

\begin{frontmatter}



\title{Addressing the Hubble tension in Yukawa cosmology?}


\author[1]{Kimet Jusufi}
\ead{kimet.jusufi@unite.edu.mk}
\affiliation[1]{organization={Physics Department, University of Tetova},   addressline={ Ilinden Street nn, 1200, Tetova, North Macedonia}}
\author[2]{Esteban González}
\ead{esteban.gonzalez@ucn.cl}
\affiliation[2]{organization={Departamento de Física, Universidad Católica del Norte},  addressline={Avenida Angamos 0610, Casilla 1280, Antofagasta, Chile}}
\author[3,4]{Genly Leon}
\ead{genly.leon@ucn.cl}
\affiliation[3]{organization={Departamento de Matemáticas, Universidad Católica del Norte},  addressline={Avenida Angamos 0610, Casilla 1280, Antofagasta, Chile}}
\affiliation[4]{organization={Institute of Systems Science, Durban University of Technology},  addressline={PO Box 1334, Durban 4000, South Africa}}

\begin{abstract}
In Yukawa cosmology, a recent discovery revealed a relationship between baryonic matter and the dark sector. The relation is described by the parameter $\alpha$ and the long-range interaction parameter $\lambda$ - an intrinsic property of the graviton. Applying the uncertainty relation to the graviton raises a compelling question: Is there a quantum mechanical limit to the measurement precision of the Hubble constant  ($H_0$)? We argue that the uncertainty relation for the graviton wavelength $\lambda$ can be used to explain a running of $H_0$ with redshift. We show that the uncertainty in time has an inverse correlation with the value of the Hubble constant. That means that the measurement of the Hubble constant is intrinsically linked to length scales (redshift) and is connected to the uncertainty in time. 
On cosmological scales, we found that the uncertainty in time is related to the look-back time quantity. For measurements with a high redshift value, there is more uncertainty in time, which leads to a smaller value for the Hubble constant. Conversely, there is less uncertainty in time for local measurements with a smaller redshift value, resulting in a higher value for the Hubble constant. Therefore, due to the uncertainty relation, the Hubble tension is believed to arise from fundamental limitations inherent in cosmological measurements. Finally, our findings indicate that the mass of the graviton fluctuates with specific scales,  suggesting a possible mass-varying mechanism for the graviton. 
\end{abstract}




\end{frontmatter}




\section{Introduction}

Based on current observations, the prevailing cosmological model depicts our Universe as homogeneous and isotropic on large scales. A particularly intriguing revelation is the existence of cold dark matter, an enigmatic form of matter that interacts solely through gravitational forces \cite{Peebles:1984zz, Bond:1984fp, Trimble:1987ee, Turner:1991id}. Despite numerous attempts, the direct detection of dark matter particles continues to be challenging, and its existence is only inferred through the gravitational effects it imparts on galaxies and broader cosmic structures \cite{Salucci:2020nlp}. 

In addition to the mysteries surrounding dark matter, dark energy has been introduced to explain the observed accelerated expansion of the Universe, as highlighted by multiple independent observations. This dark energy is closely associated with the cosmological constant \citep{Carroll:1991mt, SupernovaCosmologyProject:1997zqe, SupernovaSearchTeam:1998fmf, Peebles:2002gy}. The theoretical framework that comprehensively describes the observed cosmological phenomena is the $\Lambda$CDM paradigm, the most successful model in modern cosmology. However, fundamental questions persist regarding the nature and behavior of dark matter and dark energy \cite{Weinberg:1988cp, Zlatev:1998tr, Padmanabhan:2002ji}. A deep understanding of the Universe's physical depiction remains elusive even though scalar fields are introduced, as in the inflationary scenario \citep{Starobinsky:1980te, Guth:1980zm, DAgostino:2021vvv, DAgostino:2022fcx, Capozziello:2022tvv}. Recent inconsistencies among cosmological datasets have underscored tensions within the standard $\Lambda$CDM model, prompting questions about its accuracy in describing the entire evolution and dynamics of the universe \cite{DiValentino:2020zio, Perivolaropoulos:2021jda, DAgostino:2023cgx}. 
One of the most serious tensions is the ``Hubble tension"-- the discrepancy in $H_0$ measurements from different cosmological probes at different redshifts-- is a well-known problem in modern cosmology \cite{Vagnozzi:2019ezj, DiValentino:2019ffd, Odintsov:2020qzd, Nojiri:2021dze, Vagnozzi:2023nrq, Nojiri:2023mvi}. In particular, although the value estimated by 
the  Planck analysis   is $H_0 = (67.27\pm 0.60)$ km/s/Mpc 
\cite{Planck:2018vyg},   the direct and model-independent  measurement of the $2019$ SH0ES 
collaboration (R19) leads to $H_0 = (74.03\pm1.42)$ km/s/Mpc. 
In the literature, there is a discussion about whether the above tension is genuine or is due to possible systematics; however, currently, there seems to be a consensus that 
there is something deeper in it, possibly    a sign of new physics 
(for a complete review, the reader could see \cite{Abdalla:2022yfr}).
Nowadays, we have measurements for $H_0$ (which describes the rate at which the current Universe is expanding) that are based on observations of objects in our local Universe, say Cepheid variable stars and Type Ia supernovae (SNe Ia) \cite{Riess:2021jrx} and, on the other hand, we have the measurements on $H_0$ based on the Cosmic Microwave Background (CMB)  radiation left over from the Big Bang, which provides a snapshot of the early universe \cite{Planck:2018vyg}. The tension arises because these two methods yield different results for $H_0$. Observations from the local Universe give a higher value for the Hubble constant compared to predictions based on the CMB and the standard cosmological model. Some authors suggest that it may be possible to interpret the $H_0$-tension as a problem of unsynchronized calibrations between the early and late Universe, along with degeneracy with other parameters of our cosmological model \cite{Wagner:2022etu}. In \cite{Montani:2023xpd, Montani:2023ywn, Schiavone:2022wvq}, the authors also analyze the $H_0$ tension in modified cosmology. Regarding the oddities in observables that hint beyond $\Lambda$CDM cosmologies, with deviations at least from homogeneous and isotropic cosmologies and possibly even alternatives to General Relativity. 
Modified gravity reviews discuss cosmological issues, such as inflation, bounce, singularities, and perturbations \cite{DeFelice:2010aj, Nojiri:2010wj, Nojiri:2017ncd, Odintsov:2018uaw}.

Using the Yukawa potential in galactic systems, one can gain insights into the nature of dark matter. In this model, dark matter is postulated to be explained through the coupling between baryonic matter mediated by a long-range force represented by the Yukawa gravitational potential. The Yukawa model introduces two essential parameters: the coupling parameter $\alpha$ and the effective length parameter $\lambda$, linked to the graviton mass. The Yukawa potential finds application in various scenarios, including $f(R)$ gravity \cite{Nojiri:2003ft, Capozziello:2007ms, Nojiri:2010wj, Capozziello:2014mea, DeMartino:2017ztt, Nojiri:2017ncd, Benisty:2023ofi}. As demonstrated in recent studies \cite{Jusufi:2023xoa, Gonzalez:2023rsd}, the Yukawa potential can yield the $\Lambda$CDM as an effective model using the relation between baryonic matter ($\Omega_{B,0}$), dark energy ($\Omega_{\Lambda,0}$) and dark matter ($\Omega_{DM,0}$), i.e. $ \Omega_{DM,0}= \sqrt{2\,\Omega_{B,0}  \Omega_{\Lambda,0}}$. A similar relation has been recently obtained using the emergent nature of gravity, yielding a correspondence between emergent gravity and modified gravity \cite{Jusufi:2024rba, Jusufi:2023ayv}. In this framework, dark matter appears to have an apparent effect arising naturally from the long-range force associated with graviton, modifying Einstein's gravity at large distances. The distribution of baryonic matter undergoing the Yukawa-like gravitational interaction dictates the amount of dark matter in this perspective \cite{Capozziello:2011et, Jusufi:2023xoa, Gonzalez:2023rsd}.  
References \cite{Benisty:2022txp, Benisty:2023qcv} discusses the Yukawa potential in the galactic center and the solar system. 

In this paper, our primary focus is to investigate the $H_0$ tension within the framework of Yukawa cosmology. Specifically, our goal is to illustrate the potential connection between the $H_0$ tension and the quantum limitations of measurements associated with the graviton in cosmological contexts. We argue this is a natural consequence in Yukawa cosmology since the wavelength $\lambda$ is inherent to the graviton and plays a fundamental role in describing the energy density  $\lambda$ via $ \Omega_{\Lambda,0}= c^2\,\alpha/[\lambda^2 H^2_0 (1+\alpha)^2]$  \cite{Jusufi:2023xoa}. Possible implications of uncertainty relations in the $H_0$ tension were proposed recently in \cite{Capozziello:2020nyq, Spallicci_2022, Trivedi:2022bis}, but applied for the case of massive photons. Still, the novel idea in the present paper is that $\lambda$ is linked to the dark sector. 

The structure of the paper is as follows. Section \ref{sectII} comprehensively reviews the Yukawa potential derived from $f(R)$ gravity. Section \ref{sectIII} examines the modified Friedmann equations within the context of Yukawa cosmology, shedding light on the cosmological implications of the proposed modifications. In Section \ref{sectIV}, we address the Hubble tension. Section \ref{sectV} tests our result using observations and concludes in section \ref{sectVI}.

\section{Yukawa potential and modified gravity}
\label{sectII}
Let us briefly elaborate on the emergence of Yukawa's gravitational potential. One possible way to obtain such  Yukawa-type modifications to the Newtonian potential is to consider the weak field limit of extended theories of gravity, exemplified by $f(R)$ models \cite{Capozziello:2002rd, Sotiriou:2008rp, DeFelice:2010aj}. In these modified models of gravity, we can write down the gravitational action for $f(R)$ gravity plus matter term as
\begin{equation}
\mathcal{I}=\dfrac{1}{16\pi G}\int d^{4}x\,\sqrt{-g}\,f(R)+\mathcal{I}_{\rm matter}[g_{\mu\nu},\phi]. \label{eq.(1)}
\end{equation}

In this context, $R$ represents the Ricci scalar, $G$ denotes the Newtonian gravitational constant, and $g$ is the determinant of the metric tensor, $g_{\mu\nu}$. Notably, in the limit $f(R) \rightarrow R$, the action reverts to the Einstein-Hilbert action of general relativity. By taking variations of \eqref{eq.(1)} with respect to $g_{\mu\nu}$, we derive the field equations
\begin{equation}
f'(R)R_{\mu\nu}-\frac{1}{2}f(R)g_{\mu\nu}-f'(R)_{;\mu\nu}+g_{\mu\nu}\Box f'(R)= 8\pi G\, T_{\mu\nu},\label{eq.(2)}
\end{equation}
where $T_{\mu\nu}$ represents the matter energy-momentum tensor. The semicolon and prime symbols denote the covariant derivative and the derivative with respect to $R$, respectively, while $\Box$ represents the D'Alembert operator. Throughout this work, we use natural units of $c=\hbar=1$ unless otherwise specified.

Taking the trace of Eq. \eqref{eq.(2)}, one deduces
\begin{equation}\label{fetr}
3\Box f'(R)+f'(R)R-2f(R)= 8\pi G\, T.
\end{equation}
To illustrate the emergence of the Yukawa-like potential in $f(R)$ theories, we consider the associated field equations in the presence of matter. In the weak field limit, we can perturb the metric tensor as $g_{\mu\nu}\,=\,\eta_{\mu\nu}+h_{\mu\nu},$ where $|h_{\mu\nu}|\ll \eta_{\mu\nu}$ represents a small perturbation around the Minkowski spacetime, $\eta_{\mu\nu}$. Besides, a perturbation on the metric also acts on the Ricci scalar $R$, and then we can Taylor expand the analytic $f(R)$ about
the background value $R_0$ of $R$ \cite{Capozziello:2007ms,Capozziello:2011et,Napolitano:2012fp,Cardone:2011ze,DeMartino:2018yqf,DeLaurentis:2018ahr, Capozziello:2020dvd, Benisty:2023ofi}, say, 
\begin{align*}
f(R) &= f(R_0)+f'(R_0)(R-R_0)+f''(R_0)(R-R_0)^2/2+\hdots.
\end{align*}
In this case, $R_0=0$ is the value of $R$  for the Minkowski metric $\eta_{\mu\nu}$ in vacuum. Subsequently, imposing spherical symmetry, one can obtain $ g_{00}=-\left(1+2\Phi(r)\right)$, with the emergence of the gravitational Yukawa-like potential
\begin{equation}
\Phi(r) = -\frac{G M}{r} \left(1+\alpha \,   e^{-\frac{r}{\lambda}}\right),
\end{equation}
where $\lambda$ is the range of interaction due to the massive graviton. Let us mention here that although we refer to the massive particle as the massive graviton, it is known that $f(R)$ gravity under specific conditions can be recast in the equivalent form to scalar-tensor gravity theory. In such a case, a new degree of freedom naturally emerges as a scalar particle that couples to matter gravitationally. Therefore, we have a massless graviton as in General Relativity, plus a massive scalar particle that is sometimes known as a scalaron or, in some cases, a scalar massive graviton. The massless graviton leads to the Newtonian term, while the massive scalar graviton leads to the Yukawa correction term in the potential.  Mathematically this situation in the Yukawa potential is similar as having two gravity modes (two types of gravitons); massive and massless mode in the Yukawa potential. The Newtonian potential is obtained in the limit $\alpha=0$. The deep origin of this parameter is unknown; however, later on, we shall focus on the physical interpretation of the parameter $\alpha$. In the last equation, the rescaling has been used $G  \to  G (1+\alpha)$, along with the definitions $f'(R_0)=1+\alpha$, 
and $\lambda^2 = -\frac{1+\alpha}{6f''(R_0)}$. 
The observational constraints on the parameters of the Yukawa gravity were studied extensively in the literature. A comprehensive discussion on the value and sign of the parameter $\alpha$ to align with observations is given by~\cite{Capozziello:2007id, Capozziello:2009vr, Napolitano:2012fp, DeMartino:2018yqf}. 

Specific reasons for exploring modified gravity arise from recent findings indicating the breakdown of the Newton–Einstein law of gravity at low accelerations in wide binary systems \cite{Chae:2023prf, Chae:2023grh, Hernandez:2023qfj}. In addition, we briefly review \cite{Jusufi:2023xoa, Gonzalez:2023rsd} concerning important implications of Yukawa potential, in particular on the role of $\alpha$ and $\lambda$ on the nature of dark matter and dark energy.

The Yukawa potential can be conveniently expressed in terms of an effective length, which may be considered the wavelength of a massive graviton. In addition, one can propose a non-singular Yukawa gravitational potential of the following form \cite{Jusufi:2023xoa}
\begin{equation}
\Phi(r)=-\frac{G M m}{\sqrt{r^2+l_0^2}}\left(1+\alpha\,e^{-\frac{r}{\lambda}}\right),\; \text{where}\;   \lambda= \frac{\hbar}{m_g c},
\end{equation}
and $l_0$ is a quantum-deformed parameter and should play an important role in the Planck length domain. Utilizing the relation $F=-\nabla \Phi(r)$, we deduce the correction to Newton's law of gravitation
\begin{equation}\label{F5}
F=-\frac{G M m}{r^2}\left[1+\alpha\,\left(\frac{r+\lambda+l_0^2/r}{\lambda}\right)e^{-\frac{r}{\lambda}}\right]\left(1+\frac{l_0^2}{r^2}\right)^{-3/2}.
\end{equation}
In our discussion, for the late time universe, we shall neglect the effect of the term $l_0^2/r \to 0$. 

\subsection{Modified Friedmann equations: general case}
\label{ModifiedFLRW}
In this section, we will present the results from the study \citep{Jusufi:2023xoa} that introduces modified Friedman equations.  

Assuming the background spacetime to be spatially homogeneous and isotropic, which is given
by the Friedman-Lemaître-Robertson-Walker (FLRW) metric
\begin{equation}
ds^2=-dt^2+a^2\left[\frac{dr^2}{1-kr^2}+r^2(d\theta^2+\sin^2\theta
d\phi^2)\right], \label{FLRW}
\end{equation}
where we can further use $R=a(t)r$, $x^0=t, x^1=r$, the two dimensional metric $ h_{\mu \nu}$. 
Here $k$ denotes the
curvature of space with $k = 0, 1, -1$ corresponding to flat, closed, and open universes, respectively. The dynamical apparent
horizon, a marginally trapped surface with vanishing expansion, is
determined by the relation
\begin{equation}
h^{\mu
\nu}(\partial_{\mu}R)\,(\partial_{\nu}R)=0.
\end{equation}
We calculate the apparent horizon radius for the FLRW universe
\begin{equation}
\label{radius}
 R=ar= {1}/{\sqrt{H^2+ {k}/{a^2}}},
\end{equation}
with $H=\dot{a}/a$ being the Hubble parameter.
For the matter source in the FLRW universe, we shall assume a perfect
fluid described by the stress-energy tensor
\begin{equation}\label{T}
T_{\mu\nu}=(\rho+p)u_{\mu}u_{\nu}+pg_{\mu\nu}.
\end{equation}
On the other hand, the total mass $M = \rho V$ in the region
enclosed by the boundary $\mathcal S$ is no longer conserved, one can compute the change in the total mass using the pressure
$dM = -pdV$, and this leads to the continuity equation
\begin{equation}\label{Cont}
\dot{\rho}+3H(\rho+p)=0.
\end{equation} Let us now derive the dynamical equation for Newtonian cosmology. Toward this goal, let us consider a compact spatial region $V$ with a compact boundary
$\mathcal S$, which is a sphere having radius $R= a(t)r$, where $r$ is a dimensionless quantity. Going back and combining the second law of Newton for the test
particle $m$ near the surface, with gravitational force \eqref{F5}
we obtain
\begin{equation}\label{F6}
m\ddot{a}r=-\frac{GMm}{R^2}\left[1+\alpha\,\left(\frac{R+\lambda}{\lambda}\right)e^{-\frac{R}{\lambda}}\right] \left[1+\frac{l_0^2}{R^2}\right]^{-3/2}.
\end{equation}
We also assume $\rho=M/V$ is the energy density of the matter
inside the the volume $V=\frac{4}{3} \pi a^3 r^3$. Thus, Eq.
\eqref{F6} can be rewritten as
\begin{equation}\label{F7}
\frac{\ddot{a}}{a}=-\frac{4\pi G
}{3}\rho \left[1+\alpha\,\left(\frac{R+\lambda}{\lambda}\right)e^{-\frac{R}{\lambda}}\right] \left[1+\frac{l_0^2}{R^2}\right]^{-3/2}.
\end{equation}
This result represents the entropy-corrected dynamical equation for
Newtonian cosmology. To derive the modified Friedmann equations of the FLRW universe in general relativity, we can use the active gravitational mass $\mathcal M$ rather than the total mass $M$. It follows that, due to the
entropic corrections terms via the zero-point length, the active gravitational mass
$\mathcal M$ will be modified. Using Eq.
\eqref{F7} and replacing $M$ with $\mathcal M$, it follows 
\begin{equation}\label{M1}
\mathcal M =-\frac{\ddot{a}
a^2r^3}{G}\left[1+\alpha\,\left(\frac{R+\lambda}{\lambda}\right)e^{-\frac{R}{\lambda}}\right]^{-1} \left[1+\frac{l_0^2}{R^2}\right]^{3/2}.
\end{equation}
In addition, for the active gravitational mass, we can use the definition
\begin{equation}\label{Int}
\mathcal M =2
\int_V{dV\left(T_{\mu\nu}-\frac{1}{2}Tg_{\mu\nu}\right)u^{\mu}u^{\nu}}.
\end{equation}
From here, we show the following result
\begin{equation}\label{M2}
\mathcal M =(\rho+3p)\frac{4\pi a^3 r^3}{3}.
\end{equation}
Utilizing Eqs. \eqref{M1} and \eqref{M2} we  find
\begin{equation}\label{addot}
\frac{\ddot{a}}{a} =-\frac{4\pi G
}{3}(\rho+3p)\left[1+\alpha\,\left(\frac{R+\lambda}{\lambda}\right)e^{-\frac{R}{\lambda}}\right] \left[1-\frac{3\,l_0^2}{2\,R^2}\right].
\end{equation}
That is the modified acceleration equation for the dynamical
evolution of the  FLRW universe. To simplify the work, since $l_0$ is a very small number, we can consider a series expansion around $x=1/\lambda$ via
\begin{eqnarray}
\left[1+\alpha\,\left(\frac{R+\lambda}{\lambda}\right)e^{-\frac{R}{\lambda}}\right]=1+\alpha-\frac{1}{2}\frac{\alpha R^2}{\lambda^2}+...
\end{eqnarray}
provided that $\alpha R^2/\lambda^2 \ll1$. This relation is justified since $\alpha \ll1$, and if we take for the graviton mass $m_g \sim 10^{-68}$ kg, it implies for $\lambda \sim 10^{26} $ m, which coincides with the radius of the observable Universe $R \sim 10^{26}$ m. Otherwise, in the region, $R \rightarrow \infty$, the exponential term vanishes, i.e., $e^{-R/\lambda} \rightarrow 0$, and we get the Friedman equations in standard General Relativity \citep{Jusufi:2023xoa}. It follows that the modified Friedmann equation for $\alpha R^2/\lambda^2 \ll1$ gives
\begin{align}
\frac{\ddot{a}}{a}  =- \left(\frac{4 \pi G }{3}\right)\sum_i \left(\rho_i+3p_i\right)\left[1+\alpha-\frac{1}{2}\frac{\alpha R^2}{\lambda^2}\right]\left[1-\frac{3\,l_0^2}{2\,R^2}\right], \label{Aaddot}
\end{align}
where we have assumed several matter fluids with a constant equation of state parameters $\omega_i$ and continuity equations 
\begin{equation}\label{Cont_gen}
\dot{\rho}_i+3H(1+ \omega_i) \rho_i=0.
\end{equation}
Hence, we have the expression for densities $\rho=\rho_{i 0} a^{-3 (1+\omega_i)}$. 
Eq. \eqref{Aaddot} becomes \begin{align}
\frac{\ddot{a}}{a} =& - \left(\frac{4 \pi G }{3}\right)\sum_i \left(1+3\omega_i\right) \rho_{i 0} a^{-3 (1+\omega_i)}\nonumber \\
\times &\left[1+\alpha-\frac{1}{2}\frac{\alpha R^2}{\lambda^2}\right]\left[1-\frac{3\,l_0^2}{2\,R^2}\right]. \label{2Aaddot}
\end{align}
Next, by multiplying $2\dot{a}a$ on both sides of Eq. \eqref{2Aaddot}, we have
\begin{align}
2\dot{a}  \ddot{a} =& - \left(\frac{4 \pi G }{3}\right)\sum_i \left(1+3\omega_i\right) \rho_{i 0} a^{-3 (1+\omega_i)} 2\dot{a}a \nonumber \\
\times &\left[1+\alpha-\frac{1}{2}\frac{\alpha R^2}{\lambda^2}\right]\left[1-\frac{3\,l_0^2}{2\,R^2}\right], \\
d (\dot{a}^2+k)= & \frac{8\pi G}{3} \left[1+\alpha-\frac{1}{2}\frac{\alpha R^2}{\lambda^2}\right]\left[1-\frac{3\,l_0^2}{2\,R^2}\right]  d \left(\sum_i \rho_{i 0} a^{-1-3\omega_i)}\right), 
\end{align}
where $k$ is a constant of integration and physically characterizes the curvature of space.
Hence, with $R[a]= r a$ we have 
\begin{align}\label{Fried1}
 \dot{a}^2+k = &   \frac{8\pi G}{3} \int \left[1+\alpha-\frac{1}{2}\frac{\alpha R[a]^2}{\lambda^2}\right]\left[1-\frac{3\,l_0^2}{2\,R[a]^2}\right] \nonumber \\
\times & \frac{d \left(\sum_i \rho_{i 0} a^{-1-3\omega_i}\right)}{da} da,
\end{align}
 with $r$ nearly a constant. Hence, we have 
 \begin{align}
  & \frac{\dot{a}^2}{a^2}+\frac{k}{a^2}=  \frac{8\pi  G }{3} \left(\alpha  \left(\frac{3 l_0^2}{4\lambda ^2}+1\right)+1\right) \sum_i \rho_{i0} a^{-3 (1+\omega_{i})} \nonumber \\
  & -\frac{4 \pi  (\alpha +1) G l_0^2}{3 R^2} \sum_{i, \omega_i\neq -1}\frac{ 3  \omega_{i}+1}{\omega_{i}+1}  \rho_{i0} 
a^{-3 (1+ \omega_{i})} \nonumber \\
   & +\frac{4 \pi  \alpha  G R^2}{3 \lambda ^2}  \sum_{i, \omega_i\neq 1/3} \frac{ 1+ 3 \omega_{i}}{1-3  \omega_{i}}  \rho_{i0}  a^{-3 (1+\omega _{i})}.
 \end{align}
 In the first sum, the EoS parameter runs over all matter sources; the second sum runs over all matter sources excluding $\omega_\Lambda = -1$, and the third sum runs over all matter sources excluding radiation $\omega_R = 1/3$. That is an expected result since the last terms are ascribed to a late-time universe with negligible radiation. In contrast, the middle terms are relevant in the early Universe, where it is not expected to have constant cosmological domination.  
 
 Then, we obtain in leading order terms as $\frac{l_0^2}{\lambda ^2}\rightarrow 0$, 
\begin{align}
H^2+\frac{k}{a^2} = & \frac{8\pi G_{\rm eff}
}{3}\sum_i \rho_i -\frac{1}{R^2}  \sum_{i, \omega_i \neq -1} \Gamma_1(\omega_i)\rho_i  \nonumber \\
&  +\frac{4 \pi G_{\rm eff}}{3}R^2 \sum_{i, \omega_i\neq 1/3}\Gamma_2(\omega_i)\rho_i, \label{0Fried01}
\end{align}
where 
\begin{equation}
 R^2= \frac{a^2}{\left(a^2H^2+ {k}\right)},
\end{equation}
\begin{equation}
G_{\rm eff}=G(1+\alpha),
\end{equation}
 along with the definitions
\begin{align}
\Gamma_1 (\omega_i ) & \equiv  \frac{4 \pi G_{\rm eff} l_0^2 }{ 3 }\left(\frac{1+3
\omega_i}{1+\omega_i}\right), \quad \omega_i \neq -1\\
    \Gamma_2 (\omega_i )   & \equiv   \frac{\alpha\, (1+3\omega_i)}{  \lambda^2 (1+\alpha) (1-3\omega_i)}, \quad \omega_i\neq 1/3. \label{def-Gamma_2}
\end{align}
Observing an apparent singularity in the last equation at $\omega_R=1/3$ (radiation) adds an intriguing result to our understanding, suggesting a potential phase transition in the early Universe from a radiation-dominated state to a matter-dominated one. The modified Friedmann equation \eqref{0Fried01} was proposed in \citep{Jusufi:2023xoa} to model distinct phases of the Universe: one accounting for radiation and quantum effects during the early Universe (radiation-dominated epoch),
\begin{align}
H^2+\frac{k}{a^2} = & \frac{8\pi G_{\rm eff}
}{3}\sum_i \rho_i  -  \left(H^2+ \frac{k}{a^2}\right) \sum_{i, \omega_i \neq -1} \Gamma_1(\omega_i)\rho_i, \label{Early-Fried01}
\end{align}
and one applicable to the late-time Universe post-phase transition, characterized by a matter-dominated phase where Yukawa modifications to gravity assume a significant role,
\begin{align}
H^2+\frac{k}{a^2} = & \frac{8\pi G_{\rm eff}
}{3}\sum_i \rho_i   +\frac{4 \pi G_{\rm eff}}{3}\frac{a^2}{\left(a^2H^2+ {k}\right)} \sum_{i, \omega_i\neq 1/3}\Gamma_2(\omega_i)\rho_i. \label{Fried01}
\end{align}

Assuming only one  matter source, we obtain in leading order terms
\begin{equation}
H^2+\frac{k}{a^2} =\frac{8\pi G_{\rm eff}
}{3}\rho-\frac{\Gamma_1}{R^2}\rho+\frac{4 \pi G_{\rm eff}}{3}\rho\,\Gamma_2 R^2,
\end{equation}
where in the definitions for $\rho$, $\Gamma_1$, and $\Gamma_2$, we omitted the dependence of $\omega_i$. Using \eqref{Cont} and considering a flat universe ($k=0$), we have $R^2=1/H^2$, hence we can write 
\begin{equation}
H^2 \left(1+\Gamma_1 \rho\right)-\frac{4 \pi G_{\rm eff}}{3}\frac{\Gamma_2}{H^2}\rho=\frac{8\pi G_{\rm eff}
}{3}\rho.
\end{equation}
By expanding around $l_0$ and making use of $ \left(1+\Gamma_1 \rho\right)^{-1}\simeq \left(1-\Gamma_1 \rho\right)$ and neglecting the terms $\sim \mathcal{O}(l_0 \alpha^2/\lambda^2)$, we have
 \begin{equation}\label{imeq}
H^2-\frac{4 \pi G_{\rm eff}}{3} \frac{\Gamma_2}{ H^2} \rho  =\frac{8\pi G_{\rm eff} }{3}\rho\left(1-\Gamma_1 \rho \right).
\end{equation}

One can study two special cases for the Friedmann equation given by Eq. \eqref{imeq}. In the limit $\alpha \rightarrow 0$ [$\Gamma_2=0$], we get the quantum corrected Friedman's equations 
\begin{equation}\label{Fried2}
H^2=\frac{8\pi G }{3}\rho \left(1-\Gamma_1 \rho \right).
\end{equation}
This equation is essential to study the early Universe when the quantum effects are significant and was derived in \cite{Jusufi:2023ayv}. The phase space was further studied in \cite{Millano:2023ahb}.

On the other hand, for the late Universe, we can neglect the quantum effects and set $l_0 \rightarrow 0$ [$\Gamma_1=0$],  
 \begin{equation}\label{Gamma_2}
H^2= \frac{4 \pi G_{\rm eff}}{3} \frac{\Gamma_2}{ H^2} \rho +\frac{8\pi G_{\rm eff} }{3}\rho.
\end{equation}

Extending the model \eqref{Gamma_2} to several matter sources, we obtain 
\begin{equation}
H^2-\frac{4 \pi G_{\rm eff}}{3}\frac{1}{H^{2}}\sum_{i, \omega_i \neq 1/3} \Gamma_2(\omega_i)\,\rho_i=\frac{8\pi G_{\rm eff} }{3}\,\sum_i \rho_i,
\end{equation}
and after we use $\rho_{\rm crit}=\frac{3}{8 \pi G}H_0^2$,
we get two solutions 
\begin{align}\label{eq430}
   \frac{H^2}{H_0^2}&=\frac{(1+\alpha)}{2}\,\sum_i\Omega_i \notag \\
   &\pm  \frac{\sqrt{(\sum_i\Omega_i)^2 (1+\alpha)^2+2 \sum_{i} \frac{\Gamma_2(\omega_i) \Omega_i (1+\alpha)}{H_0^2}}}{2},
\end{align}
where  $\Omega_i=\Omega_{i0}(1+z)^{3(1+\omega_i)}, \, \Omega_{i0}=  8 \pi G \rho_{i0}/(3H_0^2)$ and the second sum under the radical runs for $\omega_i\neq 1/3$. 

\section{Yukawa cosmology and $\Lambda$CDM model }
\label{sectIII}

Now, let us consider the flat spacetime background characterized by the flat FLRW metric \eqref{FLRW} with $k=0$, such that the apparent FRW horizon radius reduces to $R=a(t)r$, where $a(t)$ denotes the normalized scale factor as a function of cosmic time.  

Therefore, using the machinery of section \ref{ModifiedFLRW}, the first Friedmann equation for a flat universe reads \citep{Jusufi:2023xoa}
\begin{equation}
H^2= \frac{8\pi G_{\rm eff}
}{3}\sum_i \rho_i +\frac{4 \pi G_{\rm eff}}{3}R^2 \sum_{i}\Gamma_2(\omega_i)\rho_i,
\label{MFried01}
\end{equation}
where $R^2=1/H^2$, $G_{\rm eff}\equiv G(1+\alpha)$ and 
$\Gamma_2 (\omega_i )$  given by \eqref{def-Gamma_2},
with $\omega_i=p_i/\rho_i$ being the equation of state parameter of the $i$-th cosmic species. Observing an apparent singularity in the last equation at $\omega_R=1/3$ (radiation) suggests a potential phase transition in the early Universe from a radiation-dominated state to a matter-dominated one. 

Summarizing, in \citep{Jusufi:2023xoa}, the revised Friedmann equations are proposed for distinct phases of the Universe: one accounting for radiation and quantum effects during the early Universe (radiation-dominated epoch), and the other applicable to the late-time Universe post-phase transition, characterized by a matter-dominated phase where Yukawa modifications to gravity assume a significant role. In what follows, we shall elaborate on two cases of Yukawa cosmology for the late-time Universe: the full model and the approximated model. 

\subsection{Full Yukawa model}

After we use $\rho_{\rm crit}=\frac{3}{8 \pi G}H_0^2$,
we get two solutions 
\begin{align}\label{eq43}
  E^2(z)&=\frac{(1+\alpha)}{2}\,\sum_i\Omega_i \notag \\
   &\pm  \frac{\sqrt{(\sum_i\Omega_i)^2 (1+\alpha)^2+2 \sum_i \Gamma_2(\omega_i) \Omega_i (1+\alpha)/H_0^2}}{2},
\end{align}
where $E(z)=H(z)/H_0, \Omega_i=\Omega_{i,0}(1+z)^{3(1+\omega_i)}$, $\Omega_{i,0}=  8 \pi G \rho_{i0}/(3H_0^2)$, and the second sum under the radical runs for $\omega_i\neq 1/3$. 

In Yukawa cosmology, we have two perfect fluid sources such that $\sum_i\Omega_i=\Omega_{B,0} 
(1+z)^{3}+\Omega_{\Lambda,0}$ in the first and second sum in Eq. \eqref{eq43}, where $z\equiv a^{-1}-1$ is the cosmological redshift, and $\Omega_{B,0}$ and $\Omega_{\Lambda,0}$ are the density parameters related to baryonic matter and dark energy, respectively. As was argued in Ref. \citep{Jusufi:2023xoa}, we consider only one source with $\omega_i=0$ in the $\Gamma_2$- corrections. Then, the physical interpretation of the term $2  \Gamma_2(0) \Omega (1+\alpha)/H_0^2 $ [where $\Gamma_2 (0)  =\frac{\alpha}{  \lambda^2 (1+\alpha)}$ and we have suppressed the subindex $i$ by defining $\Omega=\Omega_{B,0} 
(1+z)^{3}$] is closely linked to the presence of dark matter, which appears as an apparent effect in Yukawa cosmology, with fractional energy density $\Omega_{DM}$ defined through \cite{Jusufi:2023xoa}
    \begin{equation}
   \frac{\Omega^2_{DM}(1+\alpha)^2}{{(1+z)^3}}\equiv \frac{2 \Gamma_2(0) \Omega(1+\alpha)}{H_0^2}= \frac{2 \alpha \Omega_{B,0} 
(1+z)^{3}}{\lambda^2 H_0^2}.
\end{equation}
Specifically, it has been shown that the dark matter density parameter can be related to the baryonic matter as (here we reinsert the constant $c$, which was settled to $1$)
\begin{equation}\label{eqDM}
    \Omega_{DM}= \frac{c \sqrt{2 \alpha \Omega_{B,0}}}{\lambda H_0\,(1+\alpha)} 
\,{(1+z)^{3}},
\end{equation}
where the subscript `0' denotes quantities evaluated at present, specifically at $z=0$. That suggests that dark matter can be interpreted as an outcome of the modified Newton law, characterized by $\alpha$ and $\Omega_B$. Let us introduce the constant $c$ using the following definition \cite{Jusufi:2023xoa}
\begin{equation}
    \Omega_{\Lambda,0}= \frac{c^2\,\alpha}{\lambda^2 H^2_0 (1+\alpha)^2}. \label{eq:(20)}
\end{equation}
Subsequently, one can derive an expression that establishes a relation between baryonic matter, effective dark matter, and dark energy
\begin{equation}\label{DMCC}
   \Omega_{DM}(z)= \sqrt{2\,\Omega_{B,0}  \Omega_{\Lambda,0}}{(1+z)^3}\,.
\end{equation}
Considering as a physical solution only the one with the positive sign, we have 
\begin{align}\label{eq43BB}
  E^2(z)&=\frac{(1+\alpha)}{2}\,\left(\Omega_{B,0} 
(1+z)^{3}+\Omega_{\Lambda,0}\right)\notag \\
   &+\frac{(1+\alpha)}{2}\sqrt{\left(\Omega_{B,0} 
(1+z)^{3}+\Omega_{\Lambda,0}\right)^2+\frac{\Omega^2_{DM}(z)}{{(1+z)^3}}}.
\end{align}
Substituting \eqref{DMCC}
into \eqref{eq43BB}, we obtain the full Yukawa model: 
\begin{align}
 E^2(z)=\frac{1}{2} & \left(1+\sqrt{1+\frac{2\Omega_{B,0} \Omega_{\Lambda, 0} (z+1)^3}{\left(\Omega_{\Lambda, 0}+\Omega_{B,0}(z+1)^3\right)^2}}\right) \nonumber\\
   & \times (\alpha +1) \left(\Omega_{\Lambda, 0}+\Omega_{B,0} (z+1)^3\right) . \label{eq43BBB}
\end{align}

\subsection{Approximate solution: Recovering $\Lambda$CDM }
 Let us elaborate more on the late-time Universe in Yukawa cosmology. One can show how effectively the $\Lambda$CDM can be obtained by defining the order parameter
  \begin{equation}
 \delta=\frac{\Omega_{B,0} \Omega_{\Lambda, 0} (z+1)^3}{\left(\Omega_{\Lambda, 0}+\Omega_{B,0}
   (z+1)^3\right)^2}.
   \end{equation}
 Eq. \eqref{eq43BBB} can be written as 
  \begin{align}
   E^2(z)&= \frac{1}{2} \left(1+\sqrt{1+ 2 \delta}\right) (\alpha +1) \left(\Omega_{\Lambda, 0}+\Omega_{B,0} (z+1)^3\right).\label{eq43BBBB}
 \end{align}
Let us consider an expansion around $\delta=0$; 
\begin{equation}
\frac{1}{2} \left(1+\sqrt{1+ 2 \delta}\right)= 1+\frac{\delta}{2}+\mathcal{O}\left(\delta^2\right).
\end{equation}
Then, in leading-order terms, one finds 
 \begin{align}
  E^2(z)&= (\alpha +1) \left(\Omega_{\Lambda, 0}+\Omega_{B,0} (z+1)^3\right) \nonumber\\
  & \times \left(1+\frac{\Omega_{B,0} \Omega_{\Lambda, 0} (z+1)^3}{2\left(\Omega_{\Lambda, 0}+\Omega_{B,0}(z+1)^3\right)^2}\right). \label{eq43BBBBB}
\end{align}
Finally, Eq. \eqref{eq43BBBBB} can be expressed as
 \begin{equation}
   {E^2(z)}=(1+\alpha) \left[\tilde{\Omega}_{B,0} 
(1+z)^{3}+\Omega_{\Lambda,0}\right],
\end{equation}
where
\begin{equation}
  \tilde{\Omega}_{B,0}=\Omega_{B,0}\left[1+\frac{\Omega_{\Lambda,0}}{2 \left(\Omega_{\Lambda,0}+\Omega_{B,0} (1+z)^{3}\right)}\right].
\end{equation}
 As expected, the dark matter contribution is absorbed in the first $\tilde{\Omega}_{B,0}$ since we have only a baryonic matter and cosmological constant in Yukawa cosmology, and the dark matter appears only as an apparent effect as given by Eq. \eqref{eqDM} or Eq. \eqref{DMCC}. To obtain the  $\Lambda$CDM, first, we need to make the  transition from Yukawa cosmology to $\Lambda$CDM by making the replacement 
 \begin{equation}
     \left(1+\alpha\right) \tilde{\Omega}_{B,0} = \Omega_{B,0}^{\Lambda\text{CDM}}+ 
\Omega_{DM,0}^{\Lambda\text{CDM}},
 \end{equation}
 yielding \cite{Jusufi:2023xoa,Gonzalez:2023rsd}
\begin{equation}\label{HLCDM}
   {E^{2}(z)}=\left(\Omega_{B,0}^{\Lambda\text{CDM}}+\Omega_{DM,0}^{\Lambda\text{CDM}}\right)\left(1+z\right)^{3}+\Omega_{\Lambda,0}^{\Lambda\text{CDM}},
\end{equation}
where $H_{0}=H_{0}^{\Lambda\text{CDM}}$ along with the definitions
\begin{eqnarray}\notag\label{OmegasYukawa}
   \Omega_{DM,0}^{\Lambda\text{CDM}} &\equiv &\left(1+\alpha\right)\sqrt{2\,\, \Omega_{B,0} \Omega_{\Lambda,0}}=\sqrt{2 \Omega_{B,0}^{\Lambda\text{CDM}} \Omega_{\Lambda,0}^{\Lambda\text{CDM}} },\\\notag
   \Omega_{\Lambda,0}^{\Lambda\text{CDM}} &\equiv & \left(1+\alpha\right)\Omega_{\Lambda,0}={c^2 \alpha}/{[\lambda^2 H_0^2  (1+\alpha)  ]},\\
    \Omega_{B,0}^{\Lambda\text{CDM}} &\equiv & \left(1+\alpha\right)\Omega_{B,0}.
\end{eqnarray}

As discussed in the preceding section, the parameter $\alpha$ plays a crucial role in altering Newton's law. However, an intriguing question emerges regarding the fundamental origin of this parameter. As was pointed out in \cite{Jusufi:2024rba}, in some deep sense, $\alpha$ could potentially be influenced by the entanglement entropy arising from the intricate interplay between baryonic matter and the fluctuations in the gravitational field (gravitons). A subtle indication of this possibility becomes apparent when we utilize $\Omega_{\Lambda,0}^{\Lambda CDM}=\rho_{\Lambda,0}/\rho_{\rm crit}$, where $\rho_{\rm crit}=3H_0^2/(8\pi G)$ and $\rho_{\Lambda,0}=\Lambda c^2/(8 \pi G)$. From Eq. \eqref{eq:(20)} one can obtain the cosmological constant in terms of $\alpha$, as follows
\begin{equation}
    \Lambda \sim {3 \alpha }/{\lambda^2}.
\end{equation} 
In this perspective, the energy density can be interpreted as a holographic representation of dark energy, serving as an expression for the cosmological constant
\begin{equation}
    \rho_{\Lambda} \sim {3\, \alpha \,c^2}/{(8 \pi G \lambda^2)}.
\end{equation}
In the next section, we shall argue that there exists a relation between the Hubble constant and $\lambda$ via $H_0=c/\lambda$, where the role of the Hubble scale $L$  is played by $\lambda$. The emergence of the parameter $\alpha$ in our model may be attributed to the influence of entanglement entropy, primarily due to the presence of gravitons. That suggests a potential scenario where the entropy of the Universe undergoes modification by the gravitons, leading to the changes observed in gravity.

\section{Addressing the Hubble tension}
\label{sectIV}

Many ways can lead to the   alleviation of  the $H_0$ 
tension. The starting point is the relation
$\theta_s=\frac{r_s}{D_A},$
where 
$r_s\propto \int_0^{t_{recom}}dt \frac{c_s(t)}{\rho (t)}$
denotes the sound horizon while 
$D_A\propto \frac{1}{H_0} \int_{t_{recom}}^{t_{today}}dt \frac{1}{\rho (t)}$
is the angular diameter 
distance.
Hence, to eliminate the discrepancy between Planck estimation and 
late-time, model-independent measurements of $H_0$, one can try to change 
either   $r_s$ or  $D_A$ or both. Solutions that alter
the $r_s$ are often called early-time solutions. On the other hand,  
solutions changing  $D_A$ are called late-time solutions.
Concerning the second direction, one has two ways to achieve it
\cite{Heisenberg:2022gqk, Heisenberg:2022lob}. The first is to try to obtain 
a smaller effective Newton's constant since weakening gravity can, in principle, 
lead to faster expansion, namely to higher $H_0$. The second is to 
obtain extra terms in the effective dark-energy sector, which could lead to 
faster expansion, acquiring an effective dark-energy equation-of-state parameter in the phantom regime. Hence, in the literature, one can find many ways that could lead to 
late-time solutions of the $H_0$ tension \cite{DiValentino:2015ola, Hu:2015rva, Bernal:2016gxb, Kumar:2016zpg, DiValentino:2017iww, Binder:2017lkj, DiValentino:2017oaw, DiValentino:2017zyq, Sola:2017znb, Khosravi:2017hfi, Belgacem:2017cqo, DEramo:2018vss, Poulin:2018cxd, Nunes:2018xbm, DiValentino:2019jae, Vagnozzi:2019ezj, Pan:2019jqh, Pandey:2019plg, Adhikari:2019fvb, Braglia:2020auw, DAgostino:2020dhv, Barker:2020gcp, Wang:2020zfv, Ballardini:2020iws, LinaresCedeno:2020uxx, daSilva:2020bdc, Odintsov:2020qzd, Perez:2020cwa, Pan:2020bur, Benevento:2020fev, Elizalde:2020mfs, Alvarez:2020xmk, Haridasu:2020pms, DiValentino:2021izs, Adil:2021zxp, Dainotti:2021pqg, Seto:2021xua, Bernal:2021yli, Alestas:2021xes, Krishnan:2021dyb, Theodoropoulos:2021hkk, Dainotti:2022bzg, Bargiacchi:2023jse, Dainotti:2023ebr, Bargiacchi:2021hdp, Dainotti:2023bwq, Bargiacchi:2023rfd, Lenart:2022nip}.

Let us discuss in more detail some of the implications of Yukawa cosmology on the Hubble tension. In doing so, we will consider two cases: the full Yukawa cosmological model given by Eq. \eqref{eq43BBBB} and the approximated solution, which is effectively the $\Lambda$CDM, provided by Eq. \eqref{HLCDM}.

\subsection{Running Hubble tension and $\mathbb{H}_0$ diagnostic}

As a first model, we shall review a possible resolution of the Hubble tension via the running $H_0$ diagnostic recently proposed in \cite{Krishnan:2020vaf}. According to this approach, the mismatch between the Hubble constant may be achieved
through the $H_0$ diagnostic ($\mathbb{H}_0$) defined by
\begin{eqnarray}
    \mathbb{H}_0=\frac{H(z)}{\sqrt{\left(\Omega_{B,0}^{\Lambda\text{CDM}}+\Omega_{DM,0}^{\Lambda\text{CDM}}\right)\left(1+z\right)^{3}+\Omega_{\Lambda,0}^{\Lambda\text{CDM}}}}.\label{eq.(28)}
\end{eqnarray}
The above expression can explain the Hubble constant mismatch, which stems from a difference between the effective equation of state of the Universe within the FLRW cosmology framework and the current $\Lambda$CDM model. This paper will use the full Yukawa and approximated models, but it is argued that this approach generally needs to yield satisfactory results for both models.

According to \cite{Colgain:2022rxy}, all observables indicate that there is an evolution towards higher values of $\Omega_m$ between low and high redshifts, which is consistent with the expectation that they may deviate from Planck's values. However, it is essential to note that this research does not consider selection effects and other systematics across multiple observables such as SNe Ia, cosmic chronometers, BAO, and QSOs. The study supports the idea that the flat $\Lambda$CDM model is a dynamic model where the fitting parameters evolve with time, cautioning that cosmological tensions may arise from the assumption that $(H_0, \Omega_{m,0})$ are unique within the flat $\Lambda$CDM model. The result remains the same when upgrading the SNe to the Pantheon+ dataset \cite{Malekjani:2023dky}. Even after upgrading older Lyman-$\alpha$ BAO to newer Lyman-$\alpha$ BAO, the observed trends in data persist \cite{Colgain:2023bge}.

In what follows, we will give another approach to running a Hubble constant based on quantum approaches. 

\subsection{Running Hubble tension with uncertainty relations}
As noted, the $\mathbb{H}_0$ diagnostic defined by the last equation \eqref{eq.(28)} fails to provide satisfactory results when applied to the $\Lambda$CDM model. Therefore, one needs to adopt a specific equation of state for the cosmological model. Bellow, we shall give another approach to obtain a running Hubble constant with the redshift where a crucial role is played by the look-back time (and was also studied recently in \cite{Capozziello:2023ewq} concerning its' possible role on the Hubble tension). Here, the look-back time quantity can be interpreted as the uncertainty in time in cosmological observations. The wavelength $\lambda$ is intricately linked to the graviton mass. However, applying the uncertainty principle from quantum mechanics to the graviton introduces intriguing results, as we shall see. Consider the following: a large uncertainty in position corresponds to a reduced uncertainty in momentum (or graviton mass), and conversely, less uncertainty in position leads to heightened uncertainty in mass. Varied graviton mass values should naturally emerge from diverse observations due to the inherent uncertainty relation. In the case of cosmological scales, the precision of the graviton mass can be notably enhanced compared to studies focusing on galactic distances. This improvement arises from using cosmological distances, aligning with the uncertainty relation's principles. The key insight here is that the Hubble constant ($H_0$) tension may find its roots in quantum mechanical constraints inherent in cosmological measurements. Introducing new physics or other exotic explanations may not be necessary. In particular, the wavelength's dependence on specific measurements with distinct redshifts. The quantum mechanical limitations manifest differently based on the characteristics of the measurements, thereby influencing the derived properties of the graviton, such as its mass. With this in mind, let us assume for a given observation that
\begin{equation}
    \lambda (z)  \sim \hbar/(m_g c). \label{eq(27)}
\end{equation}
Hence, one can formulate the uncertainty relation in terms of position and momentum as
\begin{equation}
    \Delta x \Delta p \sim \hbar, 
\notag \text{where}\;    \lambda(z) \sim \Delta x, \; \text{and}\;   m_g c \sim \Delta p. \label{eq(28)}
\end{equation}
As the data suggests, a graviton in cosmological scales has a wavelength comparable to the size of the observable Universe, meaning such gravitons behave as a quantum mechanical object. Interestingly, Eq. \eqref{eq(27)} can be rephrased in the context of the energy-time uncertainty relation 
\begin{equation}
    \Delta E_g \Delta t \sim \hbar, \label{eq(31)} 
\end{equation}
where
\begin{equation}
    m_g c^2  \sim  \Delta E_g,
\end{equation}
where $\Delta t$ in this context gives the uncertainty in time for the graviton measured from a given distance to reach us on Earth. One can link this time to the age of the Universe $t_0$ considering the difference between this quantity, evaluated today, and that computed at a given redshift $t(z)$. Namely, we can write
\begin{equation}
    \Delta t=t_0-t(z)=t_0-a(t)t_0=z t_0/(1+z). \label{eq(33)}
\end{equation}
On the other hand, we can compute $\Delta t$ using the Hubble parameter
\begin{equation}
    H(z)=\frac{\dot{a}}{a}=\frac{1}{a}\frac{da}{dt}.   \label{eq(34)}
\end{equation}
This equation implies
\begin{equation}
    \Delta t =t_0-t(z)=\int_{t(z)}^{t_0}dt=\frac{1}{H_0} \int_0^z \frac{dz'}{(1+z')E(z')}, \label{eq(35)}
\end{equation}
where $t_0\simeq 13. 797$Gyr that provides the present values of the critical density parameters for matter and a dark 
energy component, respectively. From Eqs. \eqref{eq(33)}, \eqref{eq(34)} and \eqref{eq(35)} we obtain 
\begin{equation}
    H_0=\frac{1+z}{z t_0}\int_0^z \frac{dz'}{(1+z')E(z')}, \label{eq.(45)}
\end{equation}
where the right-hand side of \eqref{eq.(45)} is slowly varying with $z$. This relation can be, therefore, used to address the Hubble tension. The special case is obtained in the redshift limit $z\to \infty$, in that case $\Delta t=t_0$, and the Hubble constant reads
\begin{equation}
    H_0=\frac{1}{t_0}\int_0^{\infty} \frac{dz'}{(1+z')E(z')}.
\end{equation}

Introducing the dimensionless parameter \cite{Karpathopoulos:2017arc}
$\mathcal{A}= t H$, the present value of $\mathcal{A}$, denoted by $\mathcal{A}_0=t_0 H_0$, is referred to in Relativistic Cosmology as the age parameter, and it is a well-defined function in state space \cite{wainwright1997dynamical}.
In an ever-expanding model, where $a=a_0 e^{N}= a_0/(1+z)$, with $a_0=1$, denoting the current scale factor, the numbers of e-foldings $N$ assume all real values so that it can be taken the parameter of the flow $\Phi_{N}(\cdot)$. 
Let us denote by
$\mathbf{y}$ the phase space vector, such that $q$ is a phase space function.
Hence, at a fixed point $\mathbf{y}^{\star}$ of the dynamical system, $q$ is a constant, i.e., $q(\mathbf{y}^{\star})$.
Given an initial point $\mathbf{y}_0$ -- which represents our Universe in the present time, let denoted by $\mathbf{y}=\Phi_{N}(\mathbf{y}_0)$ the orbit through  $\mathbf{y}_0$ with $\Phi_0(\mathbf{y}_0)=\mathbf{y}_0$, and by
\begin{equation*}
\tilde{q}(N)=q(\Phi_{N}(\mathbf{y}_0)),
\end{equation*}
the deceleration parameter along the orbit so that $\tilde{q}(0)=q(\mathbf{y}_0)$.

Then, are deduced the expressions \cite{wainwright1997dynamical} (see also \cite{Karpathopoulos:2017arc}):
\begin{align*}
&H(N)=H_0 \exp\left[-\int_{0}^N \left(1+\tilde{q}(\tau)\right) d \tau\right],\;\text{for all}\; N\in \mathbb{R},\\
&t_0=\int_{-\infty}^{0} \frac{1}{H(N)} dN, \quad 
 t_0 H_0=\int_{-\infty}^{0} \exp\left[\int_{0}^N \left(1+\tilde{q}(\tau)\right) d \tau\right] dN,
\end{align*}
where $H_0$ is freely specifiable, this arbitrariness implies that each non-singular orbit corresponds to a 1-parameter family of physical universes conformally related by a constant rescaling of the metric. $t_0=t(0)$, denotes the value of $t$ at $\mathbf{y}_0$. The last formula implies that $\mathcal{A}_0=t_0 H_0$ is uniquely determined by the specified initial point  $\mathbf{y}_0$ on the phase space, such that $\mathcal{A}_0=t_0 H_0$ is a well-defined function on state space \cite{wainwright1997dynamical}. The present values $\Omega_i(\mathbf{y}_0)$ will restrict the location of the present state of the Universe, $\mathbf{y}_0$, in state space.

The dimensionless time $N$ has a significant geometric meaning in the physical Universe, as it is directly related to the length scale factor by the equation $a=a_0 e^N$. For any point $\mathbf{y}_0$, let $\mathbf{y}_1$ be the point on the orbit through $\mathbf{y}_0$ corresponding to the change $\Delta N$, which means $\mathbf{y}_1=\Phi_{\Delta N}(\mathbf{y}_0)$. It follows that $a_1$, the length scale factor at $\mathbf{y}_1$, is related to $a_0$ by the equation $a_1/a_0 =e^{\Delta N}$. Thus, for any point $\mathbf{y}_0$ and any change $\Delta N$, the ratio of length scales at $\mathbf{y}_0$ and $\mathbf{y}_1$  is equal to $e^{\Delta N}$. 
Any two points $\mathbf{y}_0$ and $\mathbf{y}_1$  of an ordinary orbit in the dimensionless state-space determines a unique increment $\Delta N$ in the dimensionless time variable. $\Delta N$ is the time taken for the flow to map $\mathbf{y}_0$ into $\mathbf{y}_1$. The corresponding change in the clock time $t$ is, therefore,  
\begin{equation}
  t_1 - t_0 =\frac{1}{H_0}\int_{0}^{\Delta N} \exp\left[\int_{0}^N \left(1+\tilde{q}(\tau)\right) d \tau\right] dN. \label{integral68}
\end{equation}
The integral \eqref{integral68} depends only on $\mathbf{y}_0$  and $\Delta N$ such that it can be written as 
\begin{equation}
  t_1 - t_0 =\frac{1}{H_0} \mathcal{F}(\mathbf{y}_0, \Delta N). \label{integral69}
\end{equation}
Since $H_0$ is an arbitrary positive constant, the change in the clock time $t_1 - t_0$ can be assumed to be any positive value. However, since $\mathcal{A}_0=t_0 H_0$ is determined by $\mathbf{y}_0$, it follows that the ratio $t_0/t_1$ is uniquely determined. 
As mentioned, the arbitrariness of $t_1 - t_0$ is related to an ordinary orbit corresponding to a one-parameter family of universes, all conformally related. 
To specify a unique universe evolution, one has to choose a reference point $\mathbf{y}_0$ on the orbit, which then determines $\mathcal{A}_0=t_0 H_0$ and then choose a value of $H_0$. Once it is done, two points $\mathbf{y}_0$ and $\mathbf{y}_1$ determines a unique clock time  $t_1 - t_0$ through Eq. \eqref{integral69}. 

To relate this argument to our context, we can take, say \eqref{eq43BBBB}. 
With the definition $N=-\ln(1+z)$, then, we have 
\begin{align}
   E^2(N)&=   \frac{1}{2} \left(1+\sqrt{1+ 2 \delta(N)}\right) (\alpha +1) \left(\Omega_{\Lambda, 0}+\Omega_{B,0} e^{-3 N}\right),
\end{align}
where 
 \begin{equation}
 \delta(N)=\frac{\Omega_{B,0} \Omega_{\Lambda, 0} e^{3N}}{\left(\Omega_{\Lambda, 0}e^{3N}+\Omega_{B,0}
  \right)^2} \rightarrow 0 \, \text{as}\, N\rightarrow \pm\infty.
   \end{equation}
If $\delta \ll 1$, keeping the leading terms as $N\rightarrow \pm \infty$, we have
\begin{align}
     E^2(N)&=(1+\alpha)\,\left(\Omega_{B,0} e^{-3N}+\Omega_{\Lambda,0}\right), \\ 
    & 1+  q(N) = - \frac{E'(N)}{E(N)},\\
    E(0)=1  & \implies (1+\alpha)\,\left(\Omega_{B,0} +\Omega_{\Lambda,0}\right)=1.
\end{align}
Hence, 
\begin{align}   
t_0 H_0& =\int_{-\infty}^{0} \exp\left[-\int_{0}^N  E'(\tau)/E(\tau) d \tau\right] dN  = \int_{-\infty}^{0} \frac{1}{E(N)} dN \nonumber \\
          & =\int_{-\infty}^{0} \left[(1+\alpha)\,\left(\Omega_{B,0} e^{-3N}+\Omega_{\Lambda,0}\right)\right]^{-\frac{1}{2}} dN \nonumber \\
          & =\int_{-\infty}^{0} \left[(\alpha +1) \left(1-e^{-3 N}\right) \Omega_{\Lambda, 0}+e^{-3 N}\right]^{-\frac{1}{2}} dN \nonumber \\
          & =\frac{2}{3} \frac{\tanh^{-1}\left(\sqrt{(1+\alpha)\Omega_{\Lambda,0}}\right)}{\sqrt{(1+\alpha)\Omega_{\Lambda,0} }}  \nonumber \\
          & = \frac{2}{3} \frac{\sqrt{\alpha +1} H_0 \lambda}{\sqrt{\alpha } c} \tanh
   ^{-1}\left(\frac{\sqrt{\alpha } c}{\sqrt{\alpha +1} H_0 \lambda }\right). \label{Eq66}
\end{align}
From Eq. \eqref{Eq66} it follows 
\begin{align}
  H_0 & =  \frac{\sqrt{\frac{\alpha }{\alpha +1}} c \coth
   \left(\frac{\sqrt{\frac{\alpha }{\alpha +1}} c t_0}{\lambda }\right)}{\lambda } = \frac{1}{t_0}+\frac{\alpha  c^2 t_0}{3 \lambda
   ^2}+\mathcal{O}\left(\alpha ^{3/2}\right),
\end{align}
satisfying the limit  $t_0 H_0 =1$ as $\alpha \rightarrow 0$. This relation can be, therefore, used to estimate $H_0$ in the large redshift limit. 

In the present paper, we have two cases: the full Yukawa solution where $E(z)$ is given by Eq. \eqref{eq43BBBB} and the approximated Yukawa solution where $E(z)$ is given by Eq. \eqref{HLCDM}. This expression is substituted into  Eq. \eqref{eq.(45)} to obtain the slow-varying age parameter as a function of $z$. Put differently, Eq. \eqref{eq.(45)} indicates that the Hubble constant's value is contingent on the particular measurement associated with a redshift $z$, such that we denote $H_0^{(z)}$ this slow-variation. This equation was obtained from Eq. \eqref{eq(35)}, which is also known as the look-back time and was studied recently in \cite{Capozziello:2023ewq} concerning its' possible role on the Hubble tension. In the present paper, we argued that the look-back time quantity can be interpreted as the uncertainty in time by Eq. \eqref{eq(31)} when applied to the massive graviton. The current work also demonstrates the inherent emergence of this equation within the framework of Yukawa cosmology. 

As evident from Eq. \eqref{eq(35)}, there exists an inverse relation between the uncertainty in time and the value of the Hubble constant
\begin{equation}
    \Delta t \sim \frac{z\, t_0}{1+z} \sim \frac{\kappa}{H_0^{(z)}},
\end{equation}
or, if we use Eq. \eqref{eq(27)} we can write alternatively,
\begin{equation}
    \Delta t \sim \frac{\Delta x(z)}{c} \sim \frac{\kappa}{H_0^{(z)}}, \label{eq.(40)}
\end{equation}
where $\kappa$ is some factor of proportionality. The notation $H_0^{(z)}$ shows that the observational value measured for the $H_0$ depends on the specific values of the redshift $z$. 
Regarding the last equations, a few comments are in order. For exceedingly large redshift values, such as those associated with the CMB Radiation, we encounter a scenario with greater uncertainty in time and less uncertainty in the graviton mass. Namely letting $z\gg1$, where $z+1 \sim z$, we get 
\begin{equation}
    \Delta t^{\rm CMB}= \lim_{z\gg1} \frac{z t_0}{1+z} \sim t_0 \sim \frac{\kappa^{\rm CMB}}{H_0^{\rm CMB}},
\end{equation}
we thus obtain that the uncertainty in time increases and becomes comparable to the Universe's age $\Delta t^{\rm CMB} \sim t_0$. Given that the uncertainty in time is inversely proportional to the Hubble constant, it follows that the Hubble constant will tend to have smaller values due to the inverse relation with $t_0$, i.e., $H_0^{\rm CMB} \sim \kappa^{\rm CMB}/t_0$. This results in a situation where we get a smaller value for the Hubble constant, which agrees with the $H_0=67.4 \pm 0.5$ km/s/Mpc  value reported in \cite{Planck:2018vyg}. Conversely, for smaller redshift values or local measurements, as observed in galactic scales or cases like Cepheids and SNe Ia, there is reduced uncertainty in time, leading to heightened uncertainty in the graviton mass. To see this, let us consider the case with a small value of the redshift $z$, namely 
 \begin{equation}
    \Delta t^{\rm SNe\;Ia}= \frac{z t_0}{(1+z)}  \sim \frac{\kappa^{\rm SNe\;Ia}}{H_0^{\rm SNe\;Ia}}.
\end{equation}
In this case, the uncertainty value in time $\Delta t^{\rm SNe\;Ia}$ is smaller than $t_0$, and $t_0$ is recovered only in the large limit of $z$. Therefore, for local measurements, we will have $H_0^{\rm SNe\;Ia}\sim \kappa^{\rm SNe\;Ia} (1+z)/(z t_0)$, implying that the Hubble constant should tend to have higher values compared to the large redshift limit. This effect results in a higher value for the Hubble constant, which also agrees with the $H_0=73.04 \pm  1.04$ km/s/Mpc value reported in \cite{Riess:2021jrx}.

Rewriting Eq. \eqref{eq.(40)} in terms of $\lambda$ by using $\Delta x(z) \sim \lambda (z)$, and using the rescaling $\lambda^{\rm CMB} \to \lambda^{\rm CMB}/\kappa^{\rm CMB}$, we get 
\begin{equation}
    \lambda^{\rm CMB} \sim \frac{c}{H_0^{\rm CMB}}\label{lambdacmb},
\end{equation}
and similarly $\lambda^{\rm SNe\;Ia} \to \lambda^{\rm SNe\;Ia}/\kappa^{\rm SNe\;Ia}$, gives
\begin{equation}
    \lambda^{\rm SNe\;Ia} \sim \frac{c}{H_0^{\rm SNe\;Ia}}\label{lambdaSNeIa}. 
\end{equation}
If we subtract the last two equations and we divide with $\lambda^{\rm CMB}$ it gives
\begin{equation}
    \frac{\lambda^{\rm CMB}- \lambda^{\rm SNe\;Ia} }{\lambda^{\rm CMB} }= \frac{H_0^{\rm SNe\;Ia}-H_0^{\rm CMB} }{H_0^{\rm SNe\;Ia} }.\label{eq79}
\end{equation}
The last equation shows that the Hubble tension can be linked to the different length scales in cosmology. In the general case, for any specific redshift, by differentiating, we get
\begin{equation}
    \lambda (z) \sim \frac{c}{H_0^{(z)}}  \Longrightarrow  d \lambda = \frac{c}{H_0^{(z)}} \Bigg|\frac{d H_0^{(z)}}{H_0^{(z)}}\Bigg|,\label{eq(41)}
\end{equation}
one can also get a relation between the uncertainty in the graviton's mass and the uncertainty in the Hubble constant given by
\begin{equation}
    \Delta \lambda =\frac{c}{H_0^{(z)}} \Bigg|\frac{\Delta H_0^{(z)}}{H_0^{(z)}}\Bigg|=\lambda \Bigg|\frac{\Delta H_0^{(z)}}{H_0^{(z)}}\Bigg|,
\end{equation}
where $d \lambda \simeq \Delta \lambda$, and $d H_0^{(z)} \simeq \Delta H_0^{(z)}$. Since $\lambda=\hbar/(m_g c)$, we can finally write
\begin{equation}
  \Bigg|\frac{\Delta \lambda}{\lambda}\Bigg| \sim \Bigg|\frac{\Delta m_g}{m_g}\Bigg|\sim \Bigg|\frac{\Delta H_0^{(z)}}{H_0^{(z)}}\Bigg|.
\end{equation}
A similar result was obtained for the case of massive photons in \cite{Capozziello:2020nyq}. In the present paper, we argued that this intricacy in the Hubble constant can thus be explained through the lens of fundamental limitations stemming from the uncertainty relation when applied to the graviton. The measurement of the Hubble constant is intricately tied to length scales and is intricately linked to the uncertainty in time. Hence, the challenges associated with the Hubble constant find their roots in these fundamental limitations dictated by the uncertainty relation in the context of graviton measurements. According to \eqref{eq(41)}, we can also say that the uncertainty in the graviton mass is proportional to the uncertainty in the Hubble constant. 

 Let us finally show another way of obtaining the last equation directly from the relation for dark energy density given by Eq. \eqref{eq:(20)}. Namely, we can write it down first for
\begin{equation}
    \Omega^{\rm CMB}_{\Lambda,0}= \frac{c^2\,\alpha^{\rm CMB}}{{(\lambda_0^{\rm CMB}})^2\, ({H_0^{\rm CMB}})^2 (1+\alpha^{\rm CMB})^2}.
\end{equation}
and similarly for
\begin{equation}
    \Omega^{\rm SNe\;Ia}_{\Lambda,0}= \frac{c^2\,\alpha^{\rm SNe\;Ia}}{{(\lambda_0^{\rm SNe\;Ia}})^2\, ({H_0^{\rm SNe\;Ia}})^2 (1+\alpha^{\rm SNe\;Ia})^2}.
\end{equation}
From the assumption that dark energy remains constant, i.e., $\Omega^{\rm CMB}_{\Lambda,0}= \Omega^{\rm SNe\;Ia}_{\Lambda,0}$, along with $\alpha^{\rm CMB}=\alpha^{\rm SNe\;Ia}$, we obtain the relation 
\begin{equation}
  \frac{\lambda^{\rm CMB}}{\lambda^{\rm SNe\;Ia} }=\frac{H_0^{\rm SNe\;Ia}}{H_0^{\rm CMB}}.
\end{equation}
We can subtract both sides by one; therefore, from this equation, we can easily obtain again the relation \eqref{eq79}, namely
\begin{equation}
    \frac{\lambda^{\rm CMB}}{\lambda^{\rm SNe\;Ia} }-1=\frac{H_0^{\rm SNe\;Ia}}{H_0^{\rm CMB}}-1,
\end{equation}
yielding 
\begin{equation}
    \frac{\lambda^{\rm CMB}-\lambda^{\rm SNe\;Ia}}{\lambda^{\rm SNe\;Ia} }=\frac{H_0^{\rm SNe\;Ia}-H_0^{\rm CMB}}{H_0^{\rm CMB}},
\end{equation}
then using the fact that $\lambda^{\rm SNe\;Ia}=c/H_0^{\rm SNe\;Ia}$ and $H^{\rm CMB}_0=c/\lambda^{\rm CMB}$, we get 
\begin{equation}
    \frac{\lambda^{\rm CMB}- \lambda^{\rm SNe\;Ia} }{\lambda^{\rm CMB} }= \frac{H_0^{\rm SNe\;Ia}-H_0^{\rm CMB} }{H_0^{\rm SNe\;Ia} },
\end{equation}
which coincides with our Eq. \eqref{eq79}.

\section{\label{sectV} Observational constraints}
In this section, we will study the possibility of the Yukawa cosmology in addressing the $H_{0}$ tension by confronting the full Yukawa cosmological model \eqref{eq43BBB} and their $\Lambda$CDM approximation \eqref{HLCDM} with the SNe Ia data. For that end, we compute the best-fit parameters at $1\sigma\,(68.3\%)$ of confidence level (CL) with the affine-invariant Markov chain Monte Carlo (MCMC) method \cite{Goodman_Ensemble_2010}, implemented in the pure-Python code \textit{emcee} \cite{Foreman-Mackey:2012any}. In this procedure, we have considered $100$ chains or ``walkers'' and the autocorrelation time $\tau_{\rm{corr}}$, provided by the \textit{emcee} module, as a convergence test, computing at every $50$ step the value of $\tau_{\rm{corr}}$ of each free parameter. If the current step is larger than $50\tau_{\rm{corr}}$ and the value of $\tau_{\rm{corr}}$ changed by less than $1\%$, then we will consider that the chains have converged and the constraint is stopped. This convergence test is complemented with the mean acceptance fraction (MAF), which must have a value between $0.2$ and $0.5$ \cite{Foreman-Mackey:2012any}, and can be modified by the stretch move provided by the \textit{emcee} module. For the statistics, we discard the first $5\tau_{\rm{corr}}$ steps as ``burn-in'' steps, thin by $\tau_{\rm{corr}}/2$, and we flatten the chains. Finally, since we are performing a Bayesian statistical analysis, we construct the following Gaussian likelihood function:
\begin{equation}\label{SNeLike}
    \mathcal{L}_{\rm{SNe}}\propto\exp{\left(-\frac{\chi_{SNe}^{2}}{2}\right)},
\end{equation}
where $\chi_{\rm{SNe}}^{2}$ is the merit function for the SNe Ia data, whose construction is described in subsection \ref{sec:SNe}.

\subsection{\label{sec:SNe} Type Ia supernovae data}
We will examine the SNe Ia data through the Pantheon+ sample \citep{Brout:2022vxf}, an enhanced version succeeding the original Pantheon sample \citep{Pan-STARRS1:2017jku}. This dataset comprises 1701 data points within the redshift range of $0.001\leq z\leq 2.26$. The merit function can be briefly formulated in matrix notation to facilitate our analysis, denoted by bold symbols, as:
\begin{equation}\label{meritSNe}
    \chi_{\text{SNe}}^{2}=\mathbf{\Delta D}(z,\theta,M)^{\dagger}\mathbf{C}^{-1}\mathbf{\Delta D}(z,\theta,M),
\end{equation}
with $[\mathbf{\Delta D}(z,\theta,M)]_{i}= m_{B,i}-M-\mu_{th}(z_{i},\theta)$ and $\mathbf{C}=\mathbf{C}_{\text{stat}}+\mathbf{C}_{\text{sys}}$ the total uncertainty covariance matrix, where the matrices $\textbf{C}_{\text{stat}}$ and  $\textbf{C}_{\text{sys}}$ accounts for the statistical and systematic uncertainties, respectively. In this equation, $\mu_{i}=m_{B, i}-M$ represents the observational distance modulus for the Pantheon+ sample, which is derived from a modified version of Tripp's formula \citep{Tripp:1997wt} and the use of the BBC (BEAMS with Bias Corrections) approach \citep{Kessler:2016uwi}. Consequently, the Pantheon+ sample provides the corrected apparent B-band magnitude $m_{B, i}$ of a fiducial SNe Ia at redshift $z_{i}$, with $M$ representing the fiducial magnitude of an SNe Ia. This last nuisance parameter must be performed jointly with the model's free parameters $\theta$ under investigation, for which we consider in our MCMC analysis the flat prior $-20<M<-18$. On the other hand, the theoretical distance modulus for a spatially flat FLRW spacetime is given by
\begin{equation}\label{theoreticaldistance}
    \mu_{th}(z_{i},\theta)=5\log_{10}{\left[\frac{d_{L}(z_{i},\theta)}{\text{Mpc}}\right]}+25,
\end{equation}
with $d_{L}(z_{i},\theta)$ the  luminosity distance given by
\begin{equation}\label{luminosity}
    d_{L}(z_{i},\theta)=c(1+z_{i})\int_{0}^{z_{i}}{\frac{dz'}{H_{th}(z',\theta)}},
\end{equation}
$c$ denotes the speed of light measured in units of $\text{km/s}$ and $H_{th}$ correspond to the theoretical Hubble parameter of the model confronted with observations. 

Like the Pantheon sample, a degeneracy exists between the nuisance parameter $M$ and $H_{0}$. To effectively constrain the free parameter $H_{0}$ solely through SNe Ia data, it becomes imperative to incorporate the SH0ES (Supernovae and $H_{0}$ for the Equation of State of dark energy program) Cepheid host distance anchors, as follows:
\begin{equation}\label{Cepheidmerit}
    \chi^{2}_{\text{Cepheid}}=\Delta\textbf{D}_{ 
\text{Cepheid}}\left(M\right)^{\dagger}\textbf{C}^{-1}\Delta\textbf{D}_{\text{
Cepheid}}\left(M\right),
\end{equation}
where 
$\left[\Delta\textbf{D}_{\text{Cepheid}}\left(M\right)\right]_{i}=\mu_{i}
\left(M\right)-\mu_{i}^{\text{Cepheid}}$,  with $\mu_{i}^{\text{Cepheid}}$ the 
Cepheid calibrated host-galaxy distance obtained by SH0ES \citep{Riess:2021jrx}. 
Therefore, in establishing the correspondence, Cepheid distances are utilized as the ``theory" instead of employing the model under study to calibrate $M$. This approach accounts for the sensitivity of the difference $\mu_{i}\left(M\right)-\mu_{i}^{\text{Cepheid}}$ to both $M$ and $H_{0}$, while being relatively insensitive to other parameters. In this context, the Pantheon+ sample supplies $\mu_{i}^{\text{Cepheid}}$, with the total uncertainty covariance matrix for Cepheid encapsulated within the overall uncertainty covariance matrix $\mathbf{C}$. Consequently, we can formulate the merit function for the SNe Ia data as follows:
\begin{equation}\label{SNemeritfull}
    \chi_{\text{SNe}}^{2}=\mathbf{\Delta D'}(z,\theta,M)^{\dagger}\mathbf{C}^{-1}\mathbf{\Delta D'}(z,\theta,M),
\end{equation}
where
\begin{equation}\label{SNeresidual}
    \Delta\mathbf{D'}_{i}=\left\{\begin{array}{ll}
             m_{B,i}-M-\mu_{i}^{\text{Cepheid}}, & i\in\text{Cepheid host} \\
             \\ m_{B,i}-M-\mu_{th}(z_{i},\theta), & \text{otherwise}
             \end{array}
   \right..
\end{equation}
Also, including Cepheid data is crucial to the purposes of this paper and the $H_{0}$ tension. Hence, from now on, we will exclude the consideration of the nuisance parameter $M$ from our results and concentrate solely on analyzing the free parameters specific to each model. Additionally, since the best-fit parameters minimize the merit function, $\chi_{\text{min}}^{2}$, we can utilize the evaluation of these best-fit parameters as an indicator of the goodness of the fit, where a smaller value of $\chi_{\text{min}}^{2}$ corresponds to a better fit.

The theoretical Hubble parameter is obtained from Eq. \eqref{eq43BBB} in the context of the full Yukawa cosmological model, where the condition $H(z=0)=H_{0}$ leads to
\begin{equation}\label{alphaFullYukawa}
    1+\alpha=\frac{2}{\Omega_{B,0}+\Omega_{\Lambda,0}+\sqrt{\Omega_{B,0}^{2}+\Omega_{\Lambda,0}^{2}+4\Omega_{B,0}\Omega_{\Lambda,0}}}.
\end{equation}
On the other hand, for the $\Lambda$CDM approximation of the full Yukawa model, the theoretical Hubble parameter is given by Eqs. \eqref{HLCDM} and \eqref{OmegasYukawa}, and the condition $H(z=0)=H_{0}$ leads to
\begin{equation}\label{alphaYukawa} 1+\alpha=\left[\Omega_{B,0}+\sqrt{2\Omega_{B,0}\Omega_{\Lambda,0}}+\Omega_{\Lambda,0}\right]^{-1}.
\end{equation}
Hence, the free parameters for both cosmological models are $\theta=\left\{H_{0};\Omega_{B,0};\Omega_{\Lambda,0}\right\}$, for which we consider the following flat priors: $H_{0}=100\,\frac{km/s}{Mpc}h$, with $0.55<h<0.85$, $0<\Omega_{B,0}<0.2$, and $0<\Omega_{\Lambda,0}<1$, within the condition $\alpha>0$. With these parameters, we can indirectly infer the value of $\alpha$ from Eqs. \eqref{alphaFullYukawa} and \eqref{alphaYukawa}, according to the model, $\lambda$ from Eq. \eqref{eq:(20)}, and the graviton mass from the expression $m_{g}=\hbar/(c\lambda)$. We also constrain the free parameters of the $\Lambda$CDM model as a reference model, whose Hubble parameter is given only by Eq. \eqref{HLCDM}. However, the SNe Ia data is not able to independently constraint $\Omega_{B,0}^{\Lambda\rm{CDM}}$ and $\Omega_{DM,0}^{\Lambda\rm{CDM}}$. Thus, for the $\Lambda$CDM model, we define $\Omega_{m,0}\equiv\Omega_{B,0}^{\Lambda\rm{CDM}}+\Omega_{DM,0}^{\Lambda\rm{CDM}}$, where the condition $H(z=0)=H_{0}$ leads to $\Omega_{\Lambda,0}^{\Lambda\rm{CDM}}=1-\Omega_{m,0}$, and the free parameters are $\theta=\{H_{0};\Omega_{m,0}\}$, for which we consider the flat prior $0<\Omega_{m,0}<1$. For this paper, where the constraint is made only with late-time data, it is enough to obtain a best-fit on $\Omega_{m,0}$ in the $\Lambda$CDM model. Still, for both Yukawa models, obtaining a best-fit on $\Omega_{B,0}$ is necessary because the DM is written in terms of $\Omega_{B,0}$. So, in our study, we also consider a Gaussian prior on $\Omega_{B,0}$ according to the value $\Omega_{B,0}h^{2}=0.0224\pm 0.0001$ reported in the last Planck results \cite{Planck:2018vyg}, which is weakly dependent on the cosmological model and, therefore, can be considered as a model-independent measure.

\subsection{Results and discussions}
\begin{figure}
    \centering
    \includegraphics[scale=0.34]{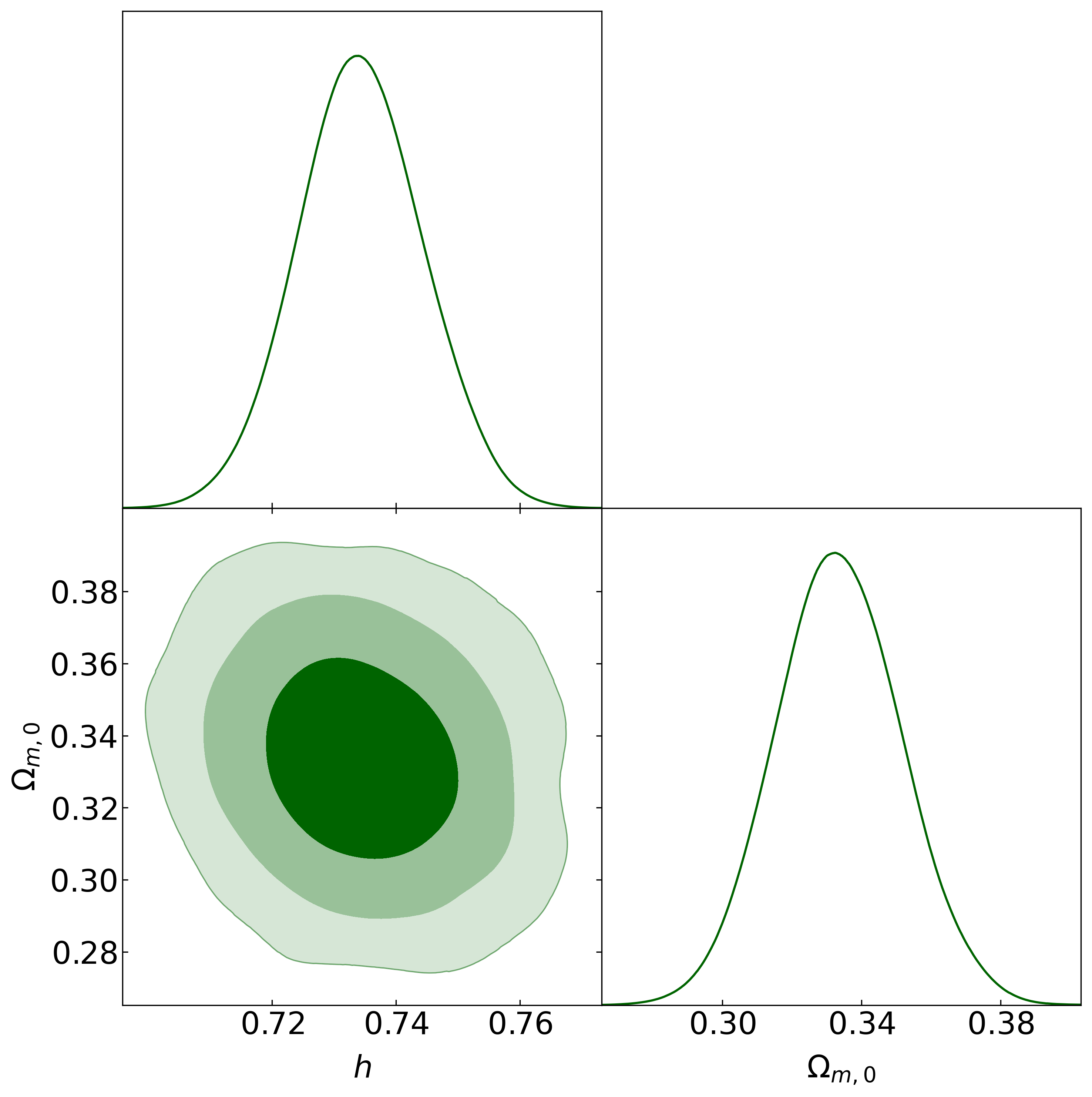}
    \caption{\label{fig:TriangleLCDM} Posterior distribution and joint marginalized regions of the free parameter space of the $\Lambda$CDM model, obtained by the MCMC analysis described in Section \ref{sectV} for the SNe Ia data. The admissible joint regions correspond to $1\sigma$, $2\sigma$, and $3\sigma$ CL, respectively. The corresponding best-fit values are shown in Table \ref{tab:best-fits}.}
\end{figure}

\begin{figure*}
	\centering
	\includegraphics[scale=0.34]{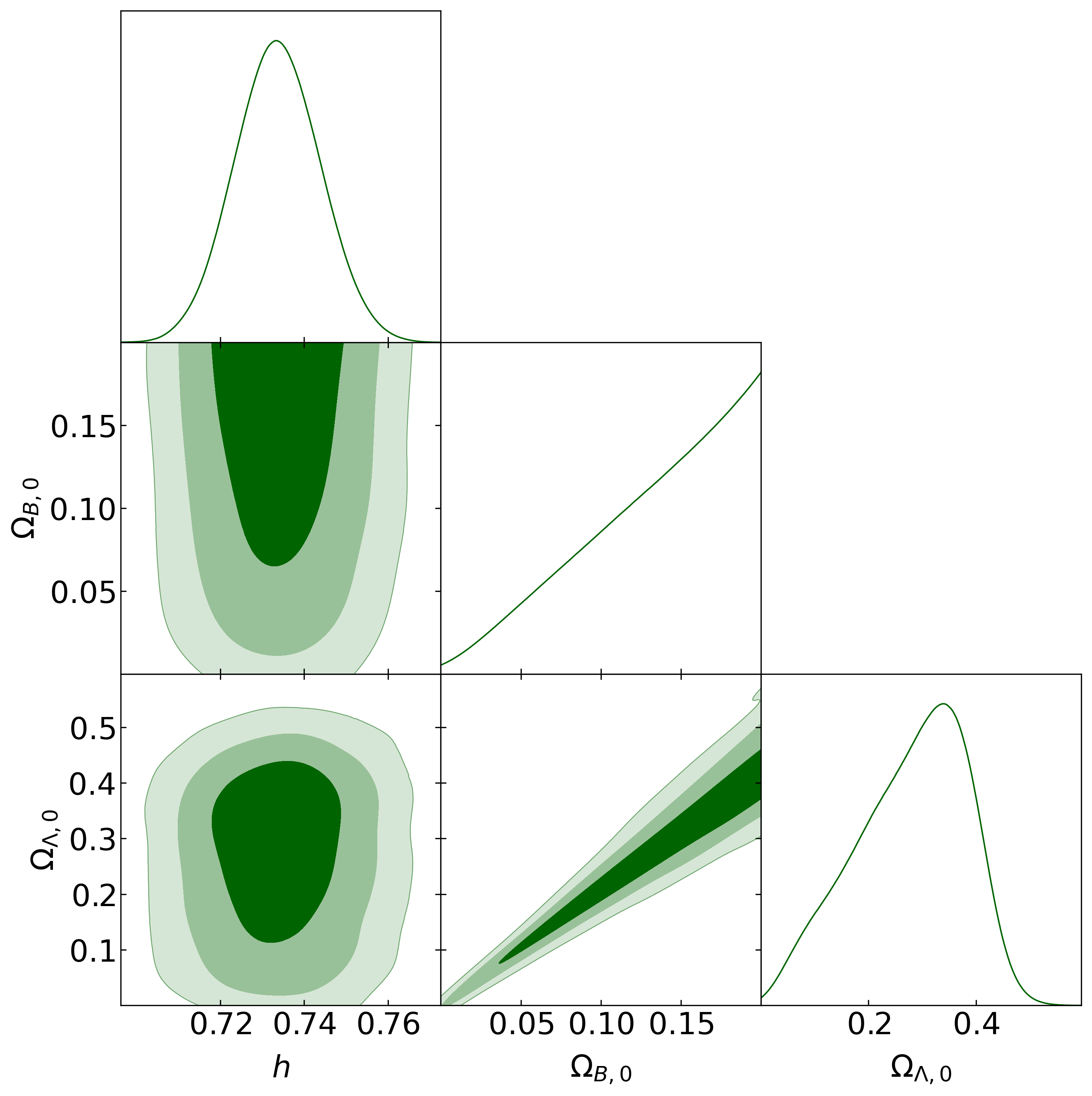}
    \includegraphics[scale=0.34]{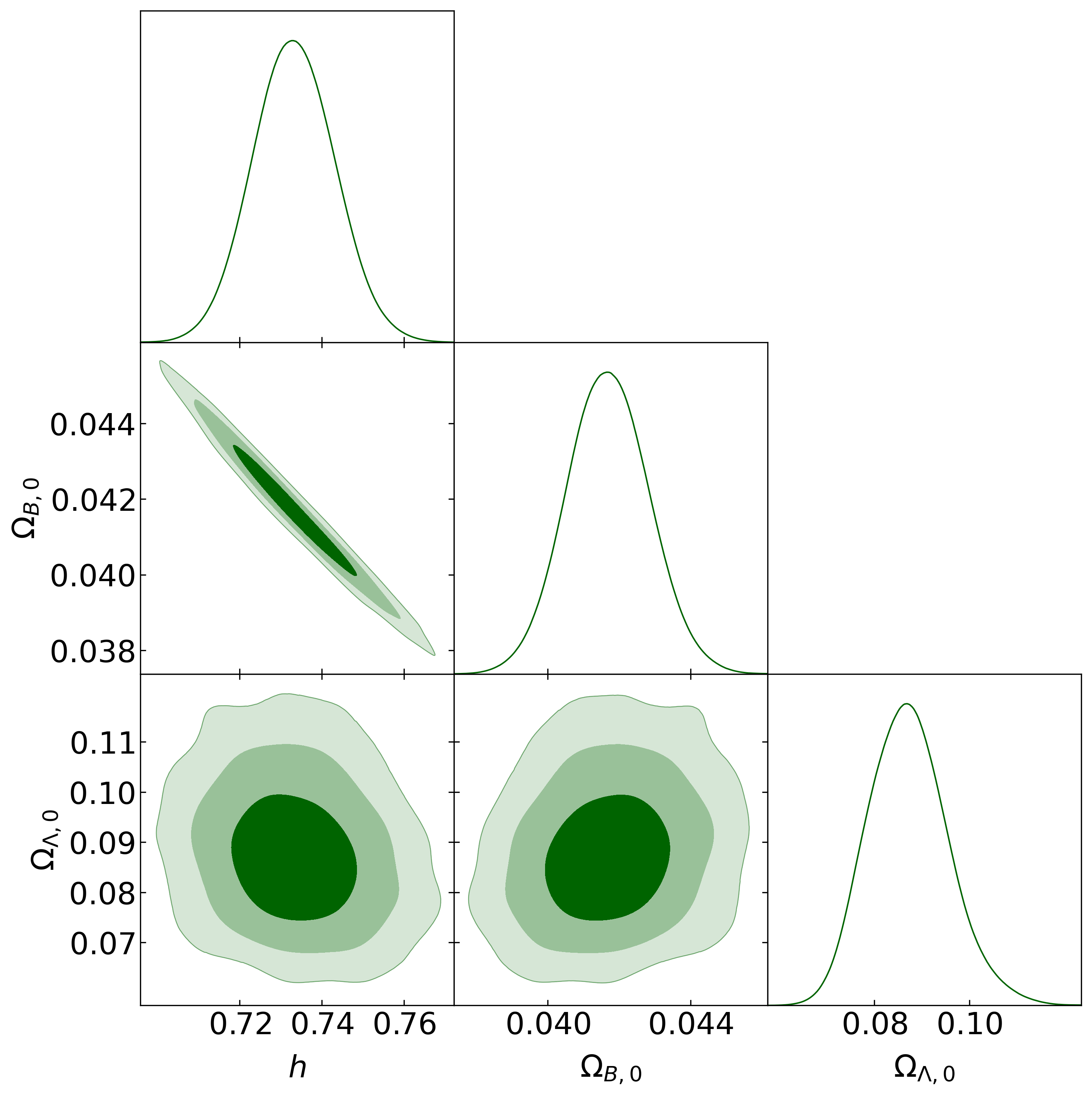}
    \caption{\label{figurefull1} Posterior distribution and joint 
    marginalized regions of the free parameters space of the full Yukawa model, obtained by the MCMC analysis described in Section \ref{sectV} for the SNe Ia data, for a flat prior on $\Omega_{B,0}$ (left panel) and a Gaussian prior on $\Omega_{B,0}$ (right panel). The admissible joint regions correspond to $1\sigma$, $2\sigma$, and $3\sigma$ CL, respectively. The corresponding best-fit values are shown in Table \ref{tab:best-fits}.}
\end{figure*}

\begin{figure*}
	\centering
	\includegraphics[scale=0.34]{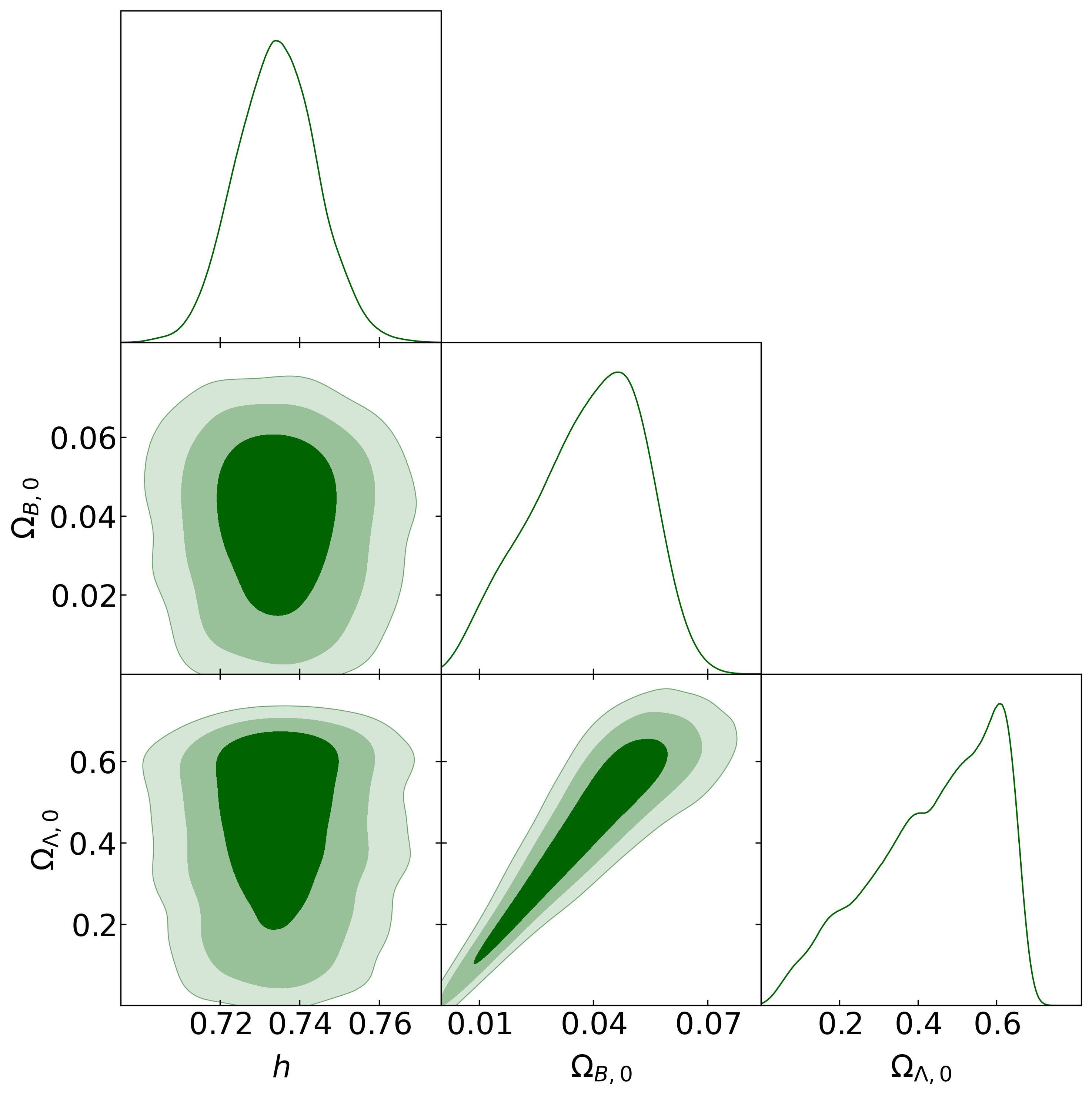}
    \includegraphics[scale=0.34]{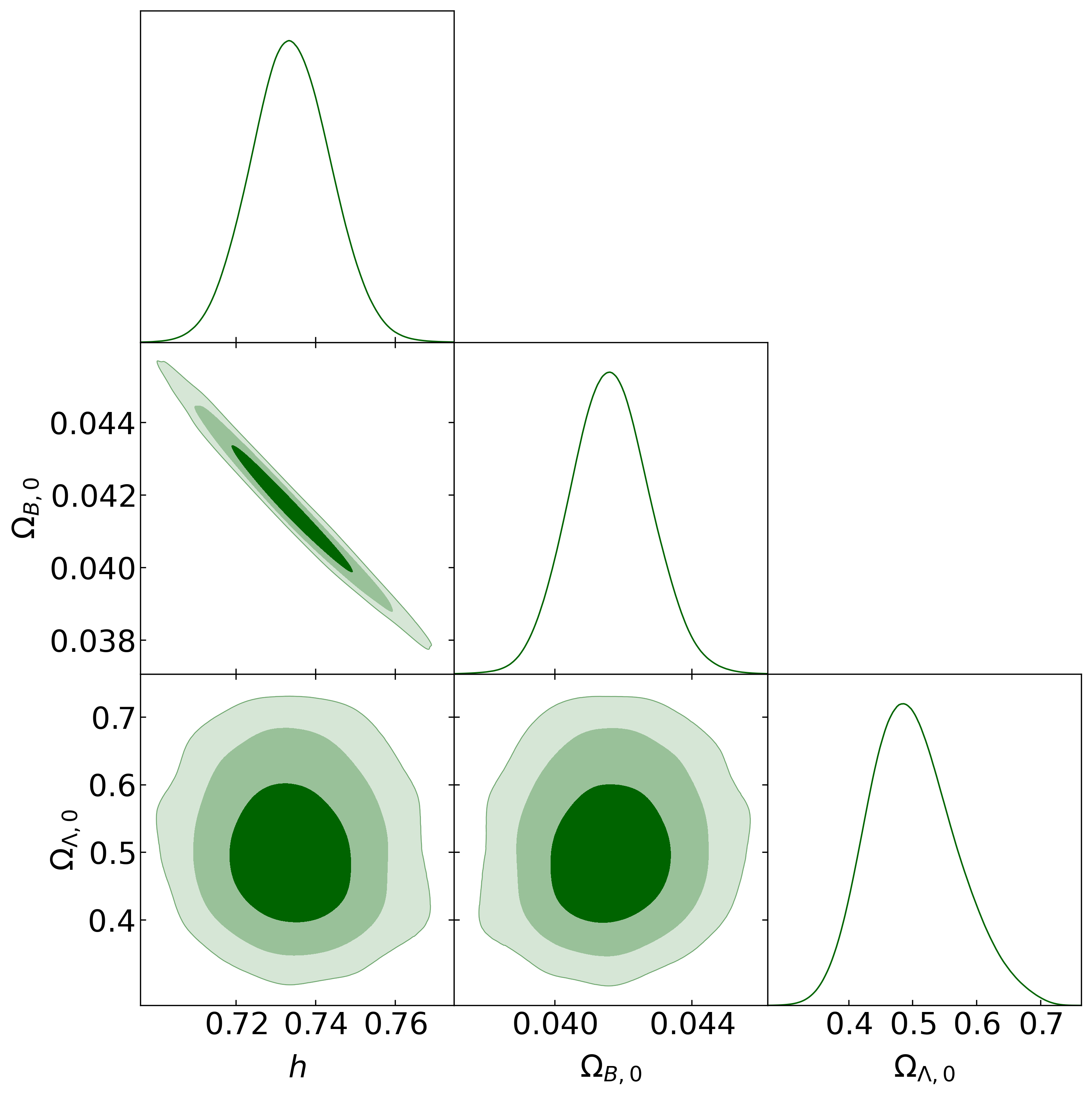}
    \caption{\label{figureapp1} Posterior distribution and joint 
    marginalized regions of the free parameters space of the approximated Yukawa model, obtained by the MCMC analysis described in Section \ref{sectV} for the SNe Ia data, for a flat prior on $\Omega_{B,0}$ (left panel) and a Gaussian prior on $\Omega_{B,0}$ (right panel). The admissible joint regions correspond to $1\sigma$, $2\sigma$, and $3\sigma$ CL, respectively. The corresponding best-fit values are shown in Table \ref{tab:best-fits}.}
\end{figure*}

In Table \ref{tab:MCMCparameters}, we present the total number of steps, the value for the stretch move, the mean acceptance fraction, and the correlation time for the free parameters space of the $\Lambda$CDM, Full Yukawa, and approximate Yukawa models. In Table \ref{tab:best-fits}, we present their respective best-fit values and $\chi_{\rm{min}}^{2}$ criteria, with the first ones at $1\sigma$ CL; while in Table \ref{tab:inferredvalues}, we present the inferred values $\alpha$, $\lambda$, and $m_{g}$ for the Yukawa models, also at $1\sigma$ CL. In Figs. \ref{fig:TriangleLCDM}, \ref{figurefull1}, and \ref{figureapp1}, we depict the posterior distribution and joint marginalized regions at $1\sigma$, $2\sigma(95.5\%)$, and $3\sigma(99.7\%)$ CL, for the free parameters space of the $\Lambda$CDM, Full Yukawa, and approximate Yukawa models, respectively. These results were obtained by the MCMC analysis described in Section \ref{sectV} for the SNe Ia data with two different priors on $\Omega_{B,0}$ for the Yukawa models.

\begin{table}
    \centering
    \begin{tabularx}{\columnwidth}{YYYYYYY}
        \hline\hline
        \multirow{2}{*}{Total steps} & \multirow{2}{*}{a} & \multirow{2}{*}{MAF} & \multicolumn{4}{c}{$\tau_{\rm{corr}}$} \\
        \cline{4-7}
         & & & $h$ & $\Omega_{m,0}$ & $\Omega_{B,0}$ & $\Omega_{\Lambda,0}$ \\
        \hline
        \multicolumn{7}{c}{$\Lambda$CDM model} \\ 
        $1250$ & $5.0$ & $0.35$ & $24.3$ & $22.7$ & $\cdots$ & $\cdots$ \\
        \hline
        \multicolumn{7}{c}{Full Yukawa model} \\
        $2600$ & $3.0$ & $0.38$ & $35.8$ & $\cdots$ & $50.9$ & $51.8$ \\
        \hline
        \multicolumn{7}{c}{Full Yukawa model Gaussian prior} \\
        $3050$ & $3.0$ & $0.42$ & $33.6$ & $\cdots$ & $27.4$ & $60.7$ \\
        \hline
        \multicolumn{7}{c}{Approximate Yukawa model} \\
        $2800$ & $3.0$ & $0.38$ & $37.2$ & $\cdots$ & $55.9$ & $55.7$ \\
         \hline
        \multicolumn{7}{c}{Approximate Yukawa model Gaussian prior} \\
        $2500$ & $3.0$ & $0.43$ & $29.7$ & $\cdots$ & $23.7$ & $31.6$ \\
        \hline\hline
    \end{tabularx}
    \caption{\label{tab:MCMCparameters} Total number of steps, value for the stretch move (a), mean acceptance fraction (MAF), and autocorrelation time $\tau_{\rm{corr}}$ for the free parameters space of the $\Lambda$CDM, Full Yukawa, and approximate Yukawa models. These values were obtained (except for $a$ which is given beforehand) when the convergence test described in Section \ref{sectV} is fulfilled for an MCMC analysis with 100 chains and the flat priors $0.55<h<0.85$, $0<\Omega_{m,0}<1$, $0<\Omega_{B,0}<0.2$, and $0<\Omega_{\Lambda,0}<1$, within the condition $\alpha>0$. Alternatively, we consider for both Yukawa models a Gaussian prior according to the value $\Omega_{B,0}h^{2}=0.0224\pm 0.0001$ reported in \cite{Planck:2018vyg}. This information is provided so that our results are replicable.}
\end{table}

\begin{table*}
    \centering
    \begin{tabularx}{\textwidth}{YYYYY}
        \hline\hline
        \multicolumn{4}{c}{Best-fit values} & \multirow{2}{*}{$\chi_{\rm{min}}^{2}$} \\
        \cline{1-4}
        $h$ & $\Omega_{m,0}$ & $\Omega_{B,0}$ & $\Omega_{\Lambda,0}$ & \\
        \hline
        \multicolumn{5}{c}{$\Lambda$CDM model}  \\
        $0.734\pm 0.010$ & $0.333\pm 0.018$ & $\cdots$ & $\cdots$ & $1523.0$ \\
        \hline
        \multicolumn{5}{c}{Full Yukawa model}  \\
        $0.734\pm 0.010$ & $\cdots$ & $0.142_{-0.063}^{+0.042}$ & $0.292_{-0.128}^{+0.092}$ & $1522.6$ \\
         \hline
        \multicolumn{5}{c}{Full Yukawa model Gaussian prior}  \\
        $0.733\pm 0.010$ & $\cdots$ & $0.042\pm 0.001$ & $0.087\pm 0.009$ & $1522.5$ \\
        \hline
        \multicolumn{5}{c}{Approximate Yukawa model}  \\
        $0.734\pm 0.010$ & $\cdots$ & $0.040_{-0.017}^{+0.013}$ & $0.470_{-0.204}^{+0.141}$ & $1523.0$ \\
         \hline
        \multicolumn{5}{c}{Approximate Yukawa model Gaussian prior}  \\
        $0.734\pm 0.010$ & $\cdots$ & $0.042\pm 0.001$ & $0.496_{-0.063}^{+0.075}$ & $1523.0$ \\
        \hline\hline
    \end{tabularx}
    \caption{\label{tab:best-fits} Best-fit values at $1\sigma$ CL and $\chi_{\rm{min}}^{2}$ criteria for the $\Lambda$CDM model, and the full and approximate Yukawa models for flat and Gaussian priors on $\Omega_{B,0}$. These values were obtained by the MCMC analysis described in Section \ref{sectV} for the SNe Ia data. Note that the Hubble parameter at the current time is written in terms of the reduced Hubble constant as $H_{0}=100\,\frac{km/s}{Mpc}h$.}
\end{table*}

\begin{table}
    \centering
    \begin{tabularx}{\columnwidth}{YYY}
        \hline\hline
        \multicolumn{3}{c}{Inferred values} \\
        \hline
        $\alpha$ & $\lambda$ [Mpc] & $m_{g}\times 10^{-42}$ [GeV] \\
        \hline
        \multicolumn{3}{c}{Full Yukawa model}  \\
        $1.09_{-0.48}^{+1.62}$ & $3772_{-582}^{+705}$ & $1.70_{-0.27}^{+0.31}$ \\
        \hline
        \multicolumn{3}{c}{Full Yukawa model Gaussian prior}  \\
        $6.07_{-0.44}^{+0.48}$ & $4833_{-121}^{+129}$ & $1.32\pm 0.03$ \\
        \hline
        \multicolumn{3}{c}{Approximate Yukawa model}  \\
        $0.41_{-0.33}^{+1.08}$ & $2707_{-1290}^{+1165}$ & $2.36_{-0.71}^{+2.15}$ \\
        \hline
        \multicolumn{3}{c}{Approximate Yukawa model Gaussian prior}  \\
        $0.35\pm 0.16$ & $2543_{-514}^{+384}$ & $2.51_{-0.33}^{+0.64}$ \\
        \hline\hline
    \end{tabularx}
    \caption{\label{tab:inferredvalues} Inferred values at $1\sigma$ CL for the full and approximate Yukawa model for flat and Gaussian priors on $\Omega_{B,0}$. These results were obtained using the chains of the best-fit values for our MCMC analysis presented in Table \ref{tab:best-fits}. $\alpha$ is computed from Eq. \eqref{alphaFullYukawa} in the case of the full Yukawa model and Eq. \eqref{alphaYukawa} in the case of the approximate Yukawa model; while $\lambda$ is computed using Eq. \eqref{eq:(20)} and the graviton mass from the expression $m_{g}=\hbar/(c\lambda)$.}
\end{table}

\begin{figure*}
	\centering
	\includegraphics[scale=0.56]{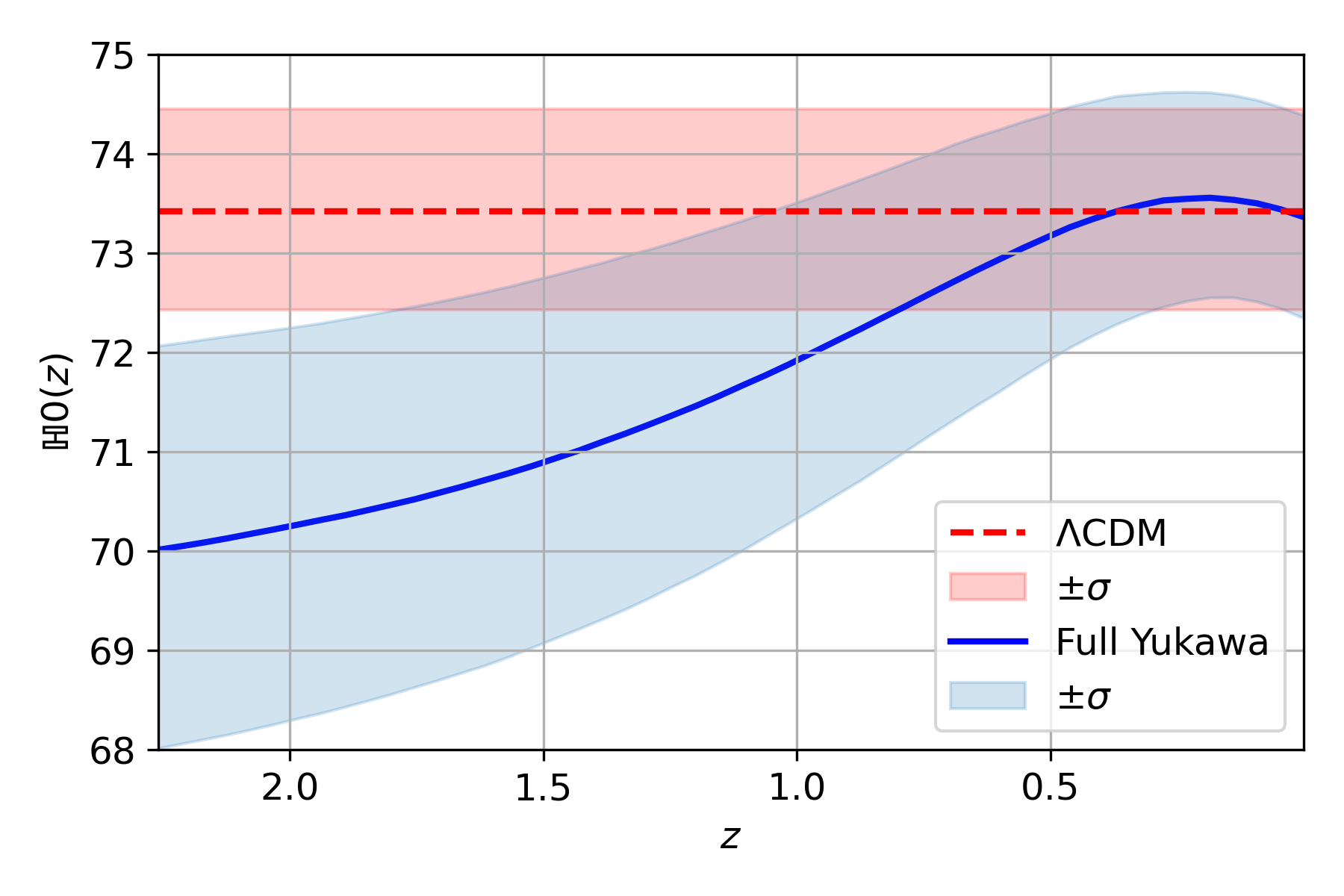}
    \includegraphics[scale=0.56]{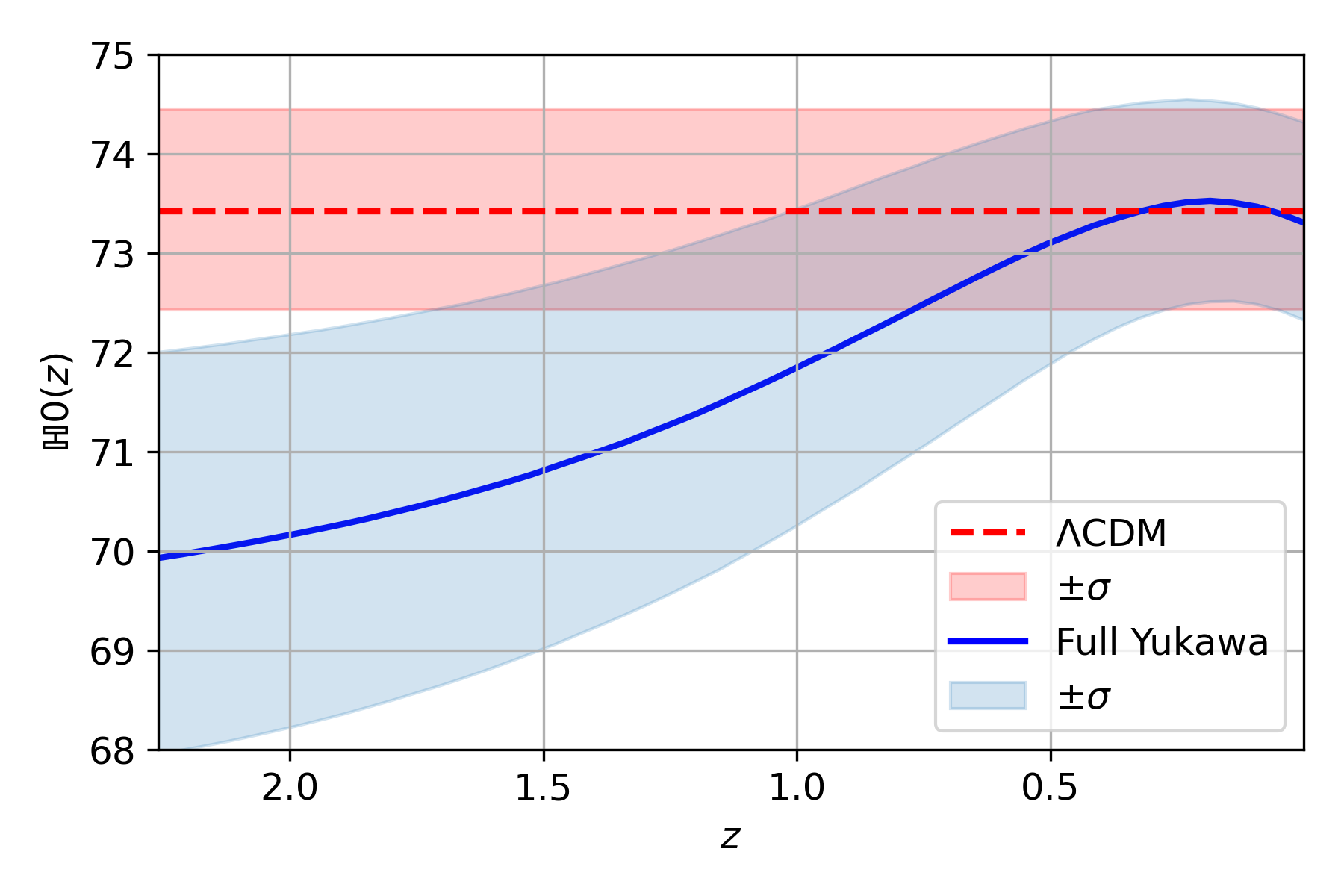}
    \caption{\label{figureH01} $\mathbb{H}_0$ diagnostic for the full Yukawa model as a function of the redshift $z$, according to Eq. \eqref{eq.(28)}, for a flat prior (left panel) and a Gaussian prior (right panel). These results were obtained using the chains of the best-fit values for our MCMC analysis presented in Table \ref{tab:best-fits}. By definition, the $\mathbb{H}_0$ diagnostic for the $\Lambda$CDM model is constant and equal to the best-fit value obtained for $H_{0}$.}
\end{figure*}

\begin{figure*}
	\centering
	\includegraphics[scale=0.56]{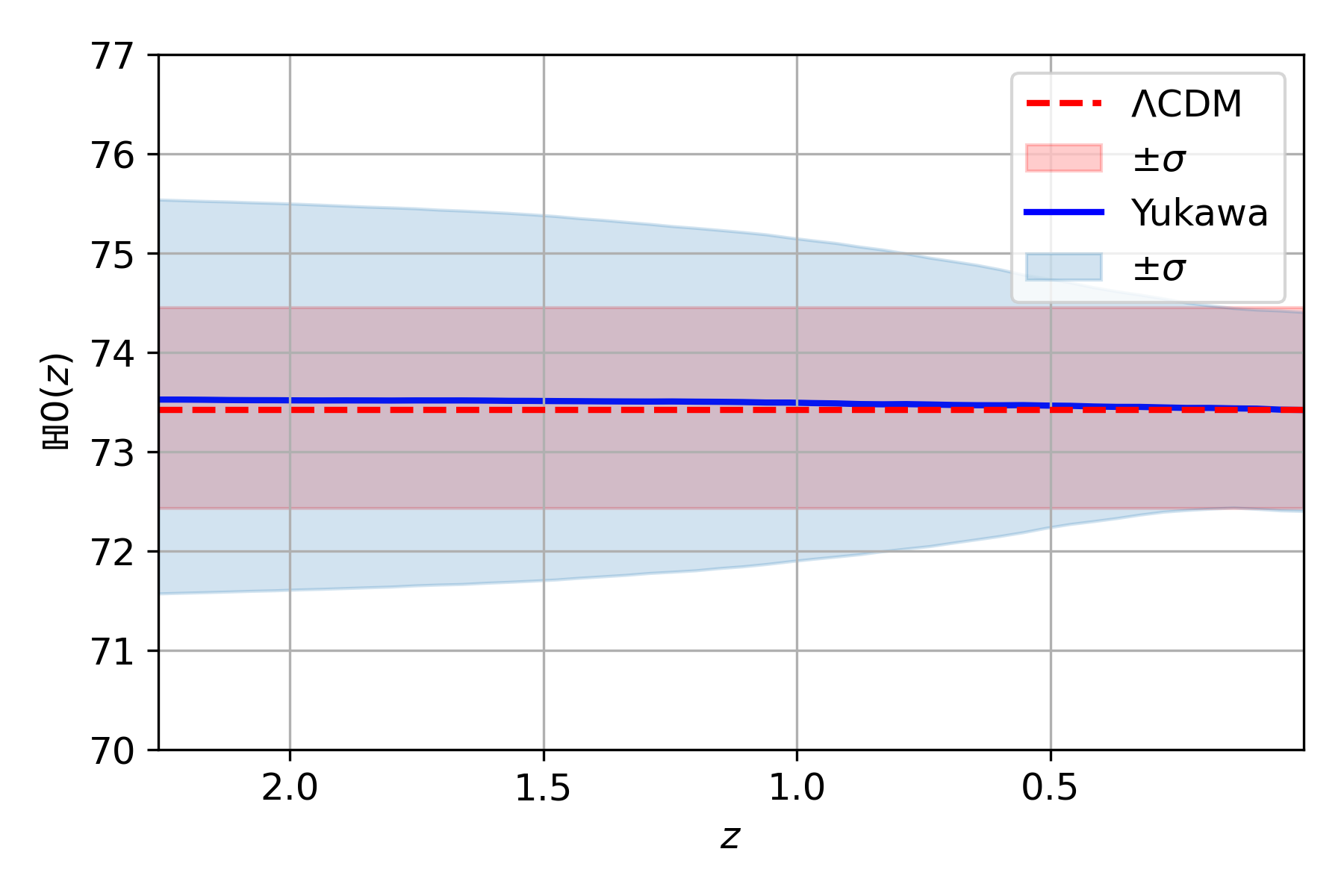}
    \includegraphics[scale=0.56]{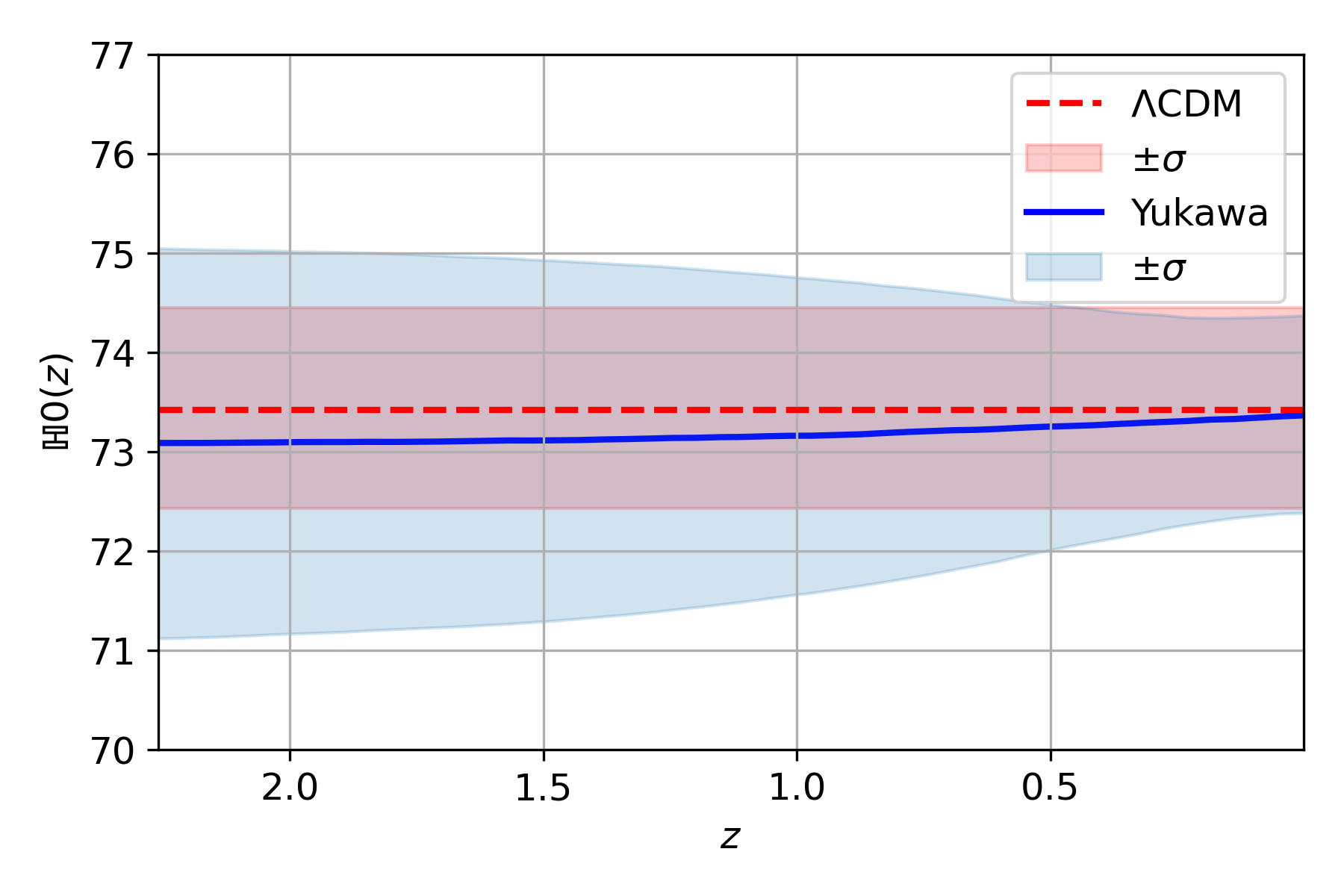}
    \caption{\label{figureH02} $\mathbb{H}_0$ diagnostic for the approximate Yukawa model as a function of the redshift $z$, according to Eq. \eqref{eq.(28)}, for a flat prior (left panel) and a Gaussian prior (right panel). These results were obtained using the chains of the best-fit values for our MCMC analysis presented in Table \ref{tab:best-fits}. By definition, the $\mathbb{H}_0$ diagnostic for the $\Lambda$CDM model is constant and equal to the best-fit value obtained for $H_{0}$.}
\end{figure*}

From Figures \ref{figurefull1} and \ref{figureapp1}, it can be seen that there is no best-fit for $\Omega_{B,0}$ at $3\sigma$ CL in the Full Yukawa model, and we have an upper bound for the approximate Yukawa model. On the other hand, at $1\sigma$ CL, there is a best-fit for $\Omega_{B,0}$ in the approximate Yukawa model and an upper bound in the Full Yukawa model. This fact is reflected in the best-fit parameters in Table \ref{tab:best-fits}. In the approximated model, the constraint on $\Omega_{B,0}$ is exclusively derived from SNe Ia data due to the relation between $\Omega_{DM,0}$ and $\Omega_{B,0}$, according to Eq. \eqref{OmegasYukawa}, effectively breaking the degeneracy between these two parameters. Additionally, the plots and the best-fits parameters reveal a difference when a Gaussian prior on $\Omega_{B,0}$ is employed, leading to improved constraints on the free parameters space of the Yukawa models, including the inferred parameters presented in Table \ref{tab:inferredvalues}. However, the values of $\chi_{min}^{2}$ remain very similar. On the contrary, in Figs. \ref{figureH01} and \ref{figureH02}, we present plots for the $\mathbb{H}_0$ diagnostic at $1\sigma$ CL as a function of the redshift $z$ for both the approximated and full Yukawa solutions, according to Eq. \eqref{eq.(28)} and using the chains of our MCMC analysis described in section \ref{sectV}. The $\mathbb{H}_0$ diagnostic indicates that the dynamical value of the Hubble constant tends to be lower than $70$ km/s/Mpc at high redshifts, potentially suggesting an alleviation of the Hubble tension. Notably, the running Hubble constant only alleviates the Hubble tension for local and large redshift values in the case of the full Yukawa model, unlike the approximated Yukawa model, which is effectively very similar to the $\Lambda$CDM model, as evident from the plots. Therefore, this approach does not yield a satisfactory result for the approximate Yukawa model.

\begin{figure*}
	\centering
    \includegraphics[scale=0.57]{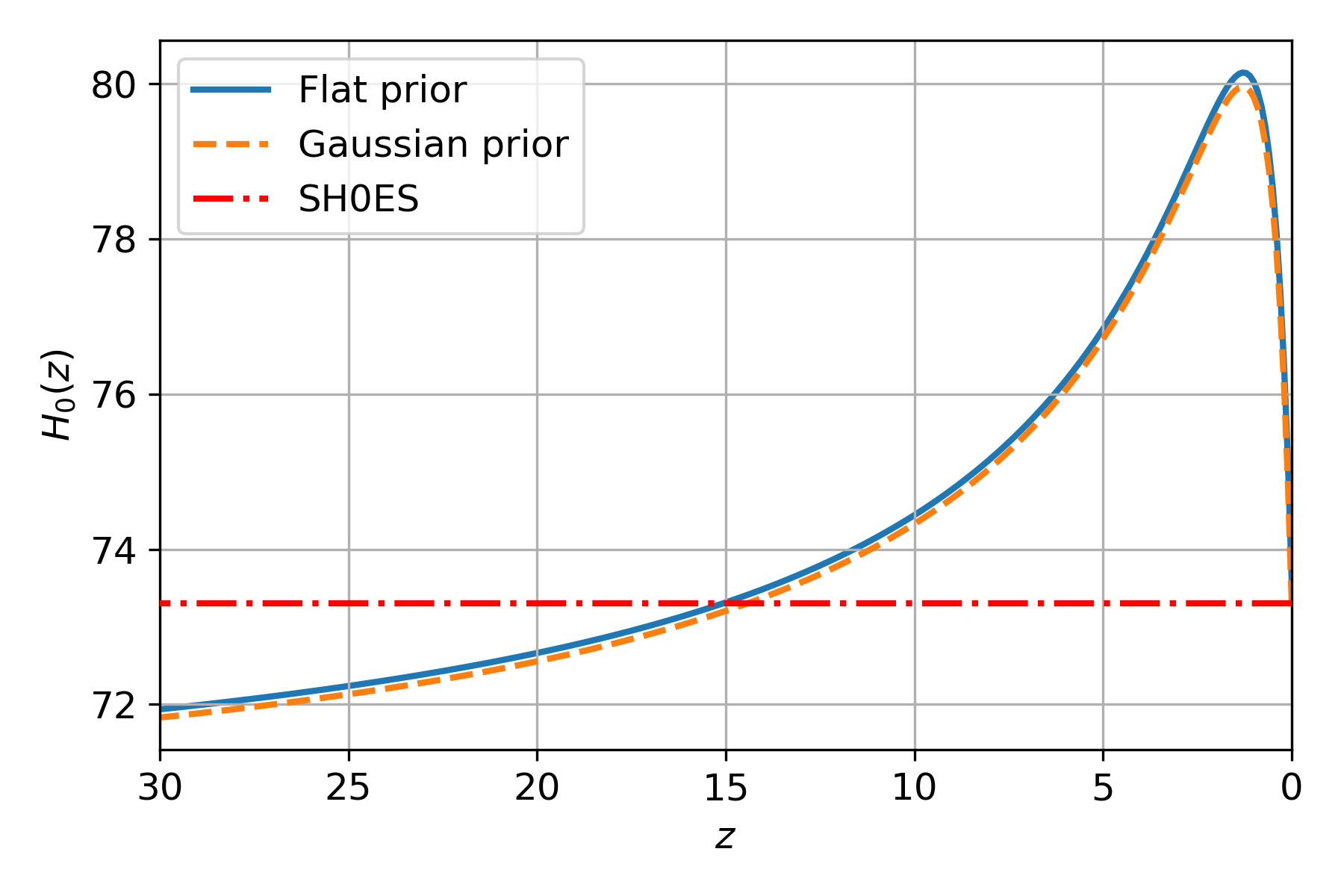}
    \includegraphics[scale=0.57]{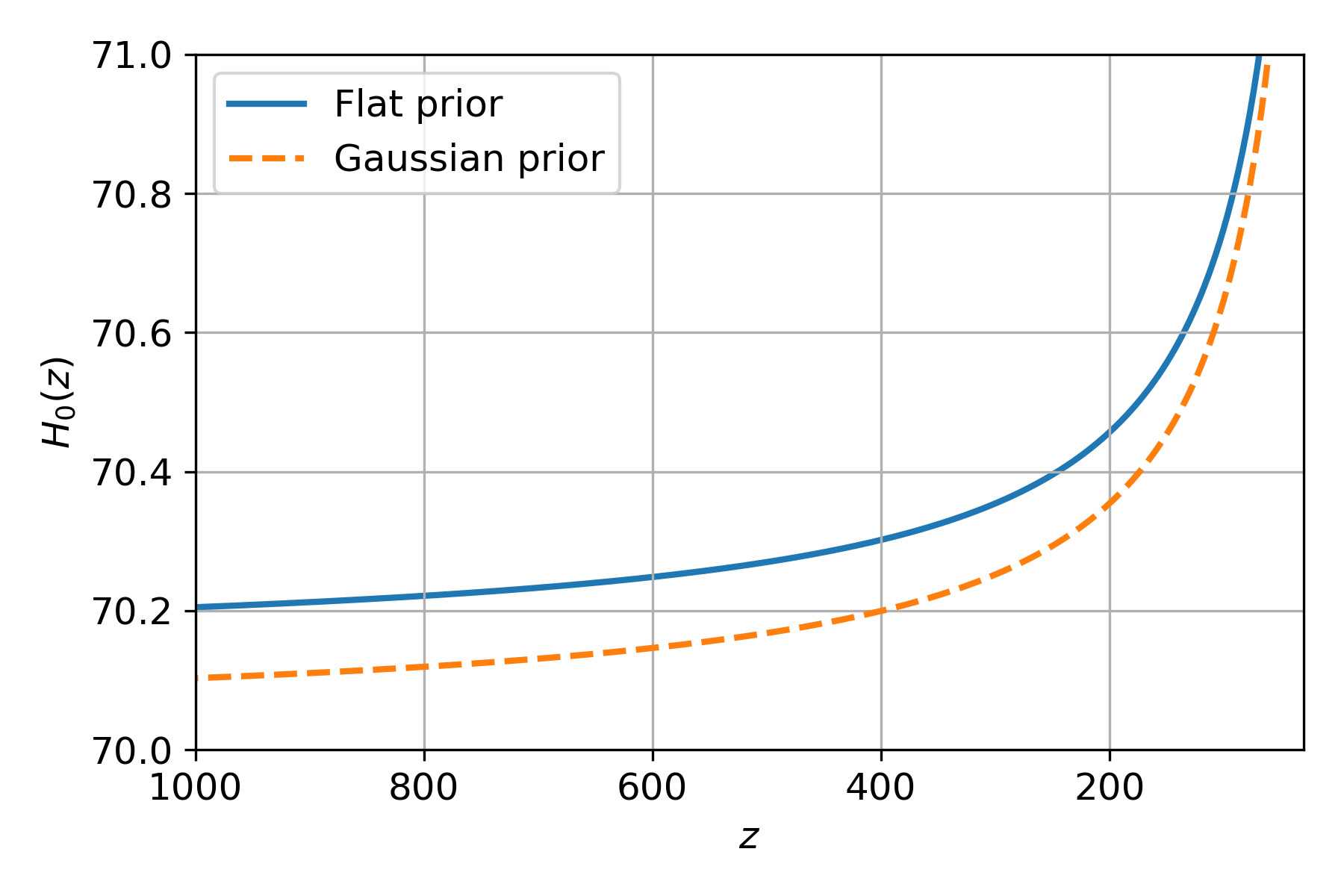}
    \caption{\label{H0full} Plot of $H_0(z)$ for the Full Yukawa model as a function of the redshift $z$, according to Eq. \eqref{eq.(45)}, in the low/high redshift interval. For the plot, we have used the best-fit values presented in Table \ref{tab:best-fits}, and we also plot in the low redshift interval the value of $H_{0}$ obtained by the SH0ES program \cite{Riess:2021jrx} for further comparison.}
\end{figure*}

\begin{figure*}
	\centering
	\includegraphics[scale=0.57]{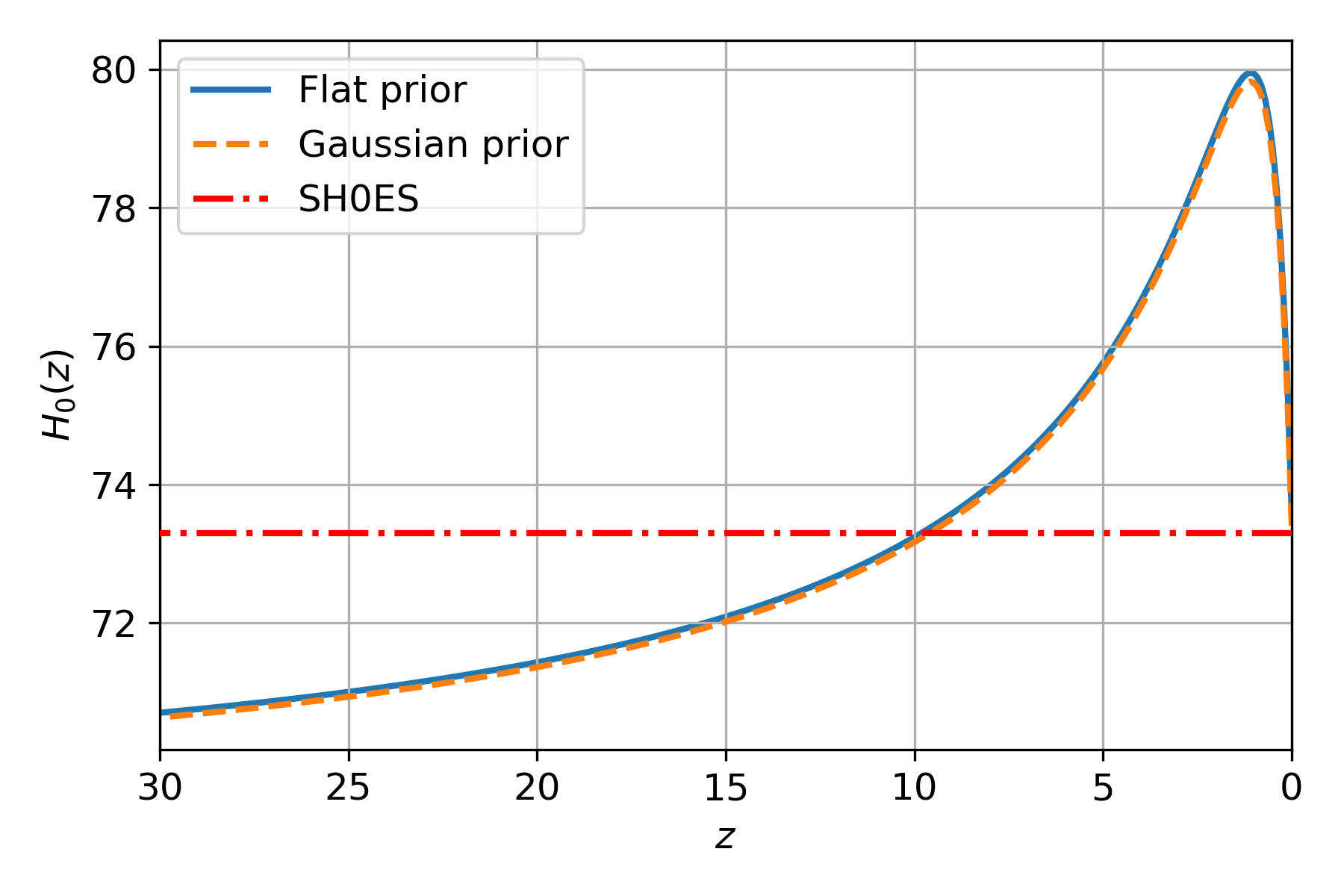}
    \includegraphics[scale=0.57]{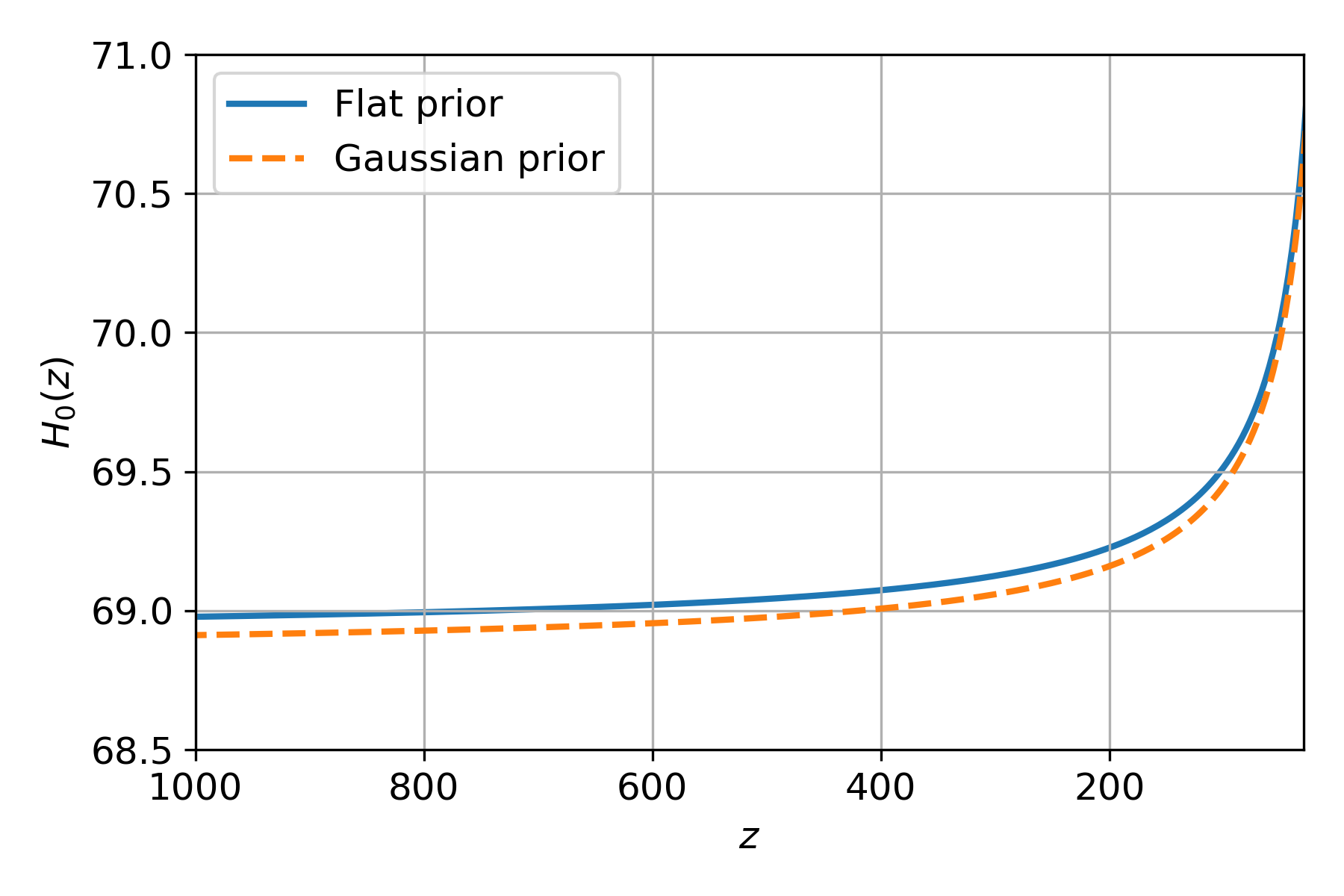}
    \caption{\label{H0appr} Plot of $H_0(z)$ for the approximate Yukawa model as a function of the redshift $z$, according to Eq. \eqref{eq.(45)}, in the low/high redshift interval. For the plot, we have used the best-fit values presented in Table \ref{tab:best-fits}, and we also plot in the low redshift interval the value of $H_{0}$ obtained by the SH0ES program \cite{Riess:2021jrx} for further comparison.}
\end{figure*}

To address the latter inconsistency, we will explore further in more detail the role of uncertainty relations on the Hubble tension. In Figs. \ref{H0full} and \ref{H0appr}, we present $H_0(z)$ as a function of the redshift $z$ according to Eq. \eqref{eq.(45)} for the Full and approximate Yukawa models, respectively. For the plot, we have used the best-fit values presented in Table \ref{tab:best-fits}, taking into account for the numerical integration the limit $H(z\to 0)\to 1/t_{0}\equiv H_{0}$. Initially, the Hubble constant increases with the increase of $z$; however, at a specific redshift, there is a turning point after which the Hubble constant decreases with increasing $z$.

As we pointed out, the Hubble tension arises from a discrepancy in the measurements of the Hubble constant $H_0$, based on observations of objects in our local Universe (Cepheid variable stars and SNe Ia) and the measurements based on the CMB radiation. As argued in the present paper, this tension might result from the measurements performed on different scales. Based on this, we showed that the following relation should hold:
\begin{equation}
  \frac{\Delta \lambda}{\lambda^{\rm CMB}} \sim \frac{\Delta m_g}{m_g^{\rm SNe\;Ia}} \sim \frac{\Delta H_0}{H_0^{\rm SNe\;Ia}}, \label{eq(44)}
\end{equation}
where 
\begin{align}
  \Delta \lambda &=  \lambda^{\rm CMB}-\lambda^{\rm SNe\;Ia},\\
  \Delta m_g &= m_g^{\rm SNe\;Ia}-m_g^{\rm CMB},\\
  \Delta H_0&= H_0^{\rm SNe\;Ia}-H_0^{\rm CMB}. \label{105}
\end{align}
Based on our analysis of the SNe Ia data, we have demonstrated that the Yukawa cosmology's Hubble constant aligns with the Hubble constant value in the $\Lambda\text{CDM}$ model. This convergence is detailed in Table \ref{tab:best-fits}. Namely, we obtained for the Hubble constant $H_0^{\rm SNe\;Ia} \sim  73.4$ km/s/Mpc. This alignment between the Hubble constant values in Yukawa cosmology and the $\Lambda\text{CDM}$ model, as demonstrated in our analysis of SNe Ia data, prompts us to extend this expectation to the analysis of CMB data. We anticipate this agreement will persist, providing further coherence between the two models. From a theoretical point of view, this fact is shown in Figs. \ref{H0full} and \ref{H0appr}, where one can see for both models that for high redshift ($z \sim 1000)$, it is possible to get a value for the Hubble constant that is close to the value obtained by Planck collaboration \cite{Planck:2018vyg}. Namely, let us take the value $H_0^{\rm CMB}=67.40 \pm 0.5$ km/s/Mpc obtained from \cite{Planck:2018vyg}, along with  $H_0^{\rm SNe\;Ia} \sim  73.4$ km/s/Mpc, and substituted in \eqref{105} to get
\begin{equation}
  {\Delta H_0}/{H_0^{\rm SNe\;Ia}} \sim 0.082. \label{eq(48)}
\end{equation}
Now, since we showed that according to Eqs. \eqref{lambdacmb} and \eqref{lambdaSNeIa}, there exists a relation between the Hubble constant and wavelength; one can compute:
\begin{equation}
     \lambda^{\rm CMB}  \sim {c}/{H_0^{\rm CMB}} \sim 4451\,{\rm Mpc},
\end{equation}
similarly, we also have, 
\begin{equation}
     \lambda^{\rm SNe\;Ia}  \sim {c}/{H_0^{\rm SNe\;Ia}} \sim 4087\, {\rm Mpc}. \label{EQ(53)}
\end{equation}
In this way, we get 
\begin{equation}
 {\Delta \lambda}/{\lambda^{\rm CMB}} \sim 0.082.
\end{equation}
Therefore, an excellent agreement exists with the result \eqref{eq(48)}. Let us, however, comment briefly on the fact that from observational constraints given in Table \ref{tab:inferredvalues}, we see that we got four values for $\lambda$ and, although all these are of Mpc order and close to the value $4087$ Mpc, they are not the same. That has to do with the fact that we have chains of data/measurements. Hence we can take as an effective value for $\lambda_{\rm eff}$ the average value, which is close to $\lambda_{\rm eff} \sim 3464$ Mpc. However, more work is needed to obtain better constraints for $\lambda$. One can also compute the corresponding mass for the graviton, but first let us see from \eqref{eq(28)} that for $\lambda^{\rm CMB}=\Delta x^{\rm CMB}$ we have more uncertainty in position but more precision on the momentum, 
\begin{equation}
    \Delta p^{\rm CMB}\sim {\hbar}/{\Delta x^{\rm CMB}} \sim 7.69 \times 10^{-61} \rm{ N\,s},
\end{equation}
compared to the case the case for $\lambda^{\rm  SNe Ia}=\Delta x^{\rm  SNe Ia}$ where we have less uncertainty in position but more uncertainty on the momentum,
\begin{equation}
    \Delta p^{\rm  SNe Ia}\sim {\hbar}/{\Delta x^{\rm  SNe Ia}}\sim 8.33 \times 10^{-61} \rm{ N\,s}. 
\end{equation}

We can further compute also the mass for each case
\begin{equation}
     m^{\rm CMB}  \sim \frac{\hbar}{\lambda^{\rm CMB} c} \sim 2.56 \times 10^{-69}  {\rm kg},
\end{equation}
and 
\begin{equation}
     m^{\rm SNe\;Ia}  \sim \frac{\hbar}{\lambda^{\rm SNe\;Ia}c} \sim  2.79 \times 10^{-69}  {\rm kg}.
\end{equation}
Finally, if we use \eqref{eq(44)} we get
\begin{equation}
{\Delta m}/{m^{\rm SNe\;Ia} } \sim 0.082.
\end{equation}

Again, as was expected, this result is in excellent agreement with Eq. \eqref{eq(48)}. The Hubble tension, therefore, might be explained by the fundamental quantum mechanical limitations of the measurements in cosmology. The relation between $\lambda$ and $H_0$ should be, in principle, universal and should hold for any measurement to add some values for $H_0$ from different measurements. Let us finally point out that the energy density of dark energy also depends on  $\lambda$, and this implies 
$\rho^{\rm SNe\;Ia}_{\Lambda} \sim 3\, \alpha \,c^2/(8 \pi G \lambda_{\rm SNe\;Ia}^2)$, which also implies fundamental quantum mechanical limitations in measuring $ \rho_{\Lambda}$. Interestingly, the uncertainty principle implies a potential dependence of the graviton's mass on cosmological scales. That suggests an underlying mechanism where varying matter densities across cosmological scales influence the mass of the graviton, akin to the chameleon mechanism \cite{Khoury:2003rn}. This idea was recently studied within the framework of massive bigravity theory \cite{DeFelice:2017oym, DeFelice:2017gzc}. If the mass of the graviton indeed varies with specific scales, it could explain observed fluctuations, such as those seen in the CMB and SNe Ia.

\section{Conclusions}
\label{sectVI}
In the present paper, we used the Yukawa modified gravity, which has two parameters, $\alpha$ and $\lambda$, related to the graviton mass, to address the Hubble tension. In particular, we used two models: full and approximated Yukawa models. To address the Hubble tension, first, we used the $\mathbb{H}_0$ diagnostic Hubble constant, a showdown that it tends to be lower than $70$ km/s/Mpc at high redshifts,  suggesting a possible alleviation of the Hubble tension. However, the running Hubble constant only tends to alleviate the Hubble tension in the case of the full Yukawa model, unlike the approximated Yukawa model, which is effectively very similar to the $\Lambda$CDM model, indicating that the resolution of the Hubble tension could be more attainable through our approach.

This paper primarily investigates the impact of uncertainty relations on the phenomenon of Hubble tension, constituting its main result. Since dark matter and dark energy are related to each other in terms of gravitons' wavelength $\lambda$, it is natural to apply the uncertainty relations to the graviton where $\lambda$ is defined as $\lambda=\hbar/(m_g c)$. In doing so, we have obtained the energy-time uncertainty relation, which shows that the uncertainty in time exhibits an inverse correlation with the value of the Hubble constant, meaning that the measurement of the Hubble constant is intricately tied to length scales and to the uncertainty in time. The MCMC analyses using the SNe Ia data showed that the Hubble constant in Yukawa cosmology is the same as in $\Lambda$CDM. In general, this is expected to be the case for CMB data. As we have argued in the present paper,  to resolve this inconsistency, in cosmological scales, it was found that the uncertainty in time coincides with the look-back time quantity. Therefore, the Hubble constant, in general, depends on the measurements of the length scale with a given redshift. That indicates that we have a smaller value for the Hubble constant when there is increased uncertainty in time, especially for measurements with a high redshift value (such as the CMB). Similarly, if there is less uncertainty in time, the case for local measurements with a smaller redshift value (such as the Cepheids and SNe Ia) implies a higher value for the Hubble constant. Interestingly enough, it is also shown how the same  conclusions emerge directly from the relation of the dark energy density parameter $\Omega_{\Lambda,0}$ in Yukawa cosmology with the assumption that this quantity should remain constant along with the coupling parameter $\alpha$.  These results show that the Hubble tension arises due to fundamental limitations inherent in cosmology measurements of the graviton's wavelength (scales) by the uncertainty relation.
 
\bigskip

\section*{CRediT authorship contribution statement}
\textbf{Kimet Jusufi:} Conceptualization, Methodology, Formal analysis, Investigation, Writing – original draft, Writing – review \& editing, Supervision.
\textbf{Esteban González:} Conceptualization, Methodology, Software, Visualization, Validation, Formal analysis, Investigation, Writing – review \& editing, Funding acquisition, Supervision, Project administration. \textbf{Genly Leon:} Conceptualization, Methodology, Formal analysis, Investigation, Writing – review \& editing, Funding acquisition, Supervision, Project administration. 

\section*{Declaration of competing interest}
The authors declare that they have no known competing financial interests or personal relations that could have appeared to influence the work reported in this paper.

\section*{Acknowledgments}
E. G. was funded by Vicerrectoría de Investigación y Desarrollo Tecnológico (VRIDT) at Universidad Católica del Norte (UCN) through Resolución VRIDT N°076/2023. He also acknowledges the scientific support of Núcleo de Investigación No. 7 UCN-VRIDT 076/2020, Núcleo de Modelación y Simulación Científica (NMSC). G. L. acknowledges the financial support of ANID through Proyecto Fondecyt Regular 2024,  Folio 1240514, Etapa 2024, Resolución VRIDT No. 026/2023, Resolución VRIDT No. 027/2023,  Proyecto de Investigación Pro Fondecyt Regular 2023 (Resolución VRIDT N°076/2023) and Resolución VRIDT N°09/2024. He also acknowledges the scientific support of Núcleo de Investigación Geometría Diferencial y Aplicaciones (Resolución VRIDT No. 096/2022).

\section*{Data Availability}
Section \ref{sectV} cited the data underlying this article.

\bibliographystyle{apsrev4-1}
\bibliography{references.bib}

\end{document}